\RequirePackage[l2tabu, orthodox]{nag}
\documentclass[aps, pra, showpacs, superscriptaddress, longbibliography, twocolumn]{revtex4-2}
\pdfoutput=1
\usepackage[dvips]{graphics,color}
\usepackage{amssymb}
\usepackage{amsmath}
\usepackage{amsfonts}
\usepackage{mathtools}
\usepackage{graphicx}
\usepackage[colorlinks,
            linkcolor=blue,
            anchorcolor=blue,
            citecolor=blue,
            urlcolor=blue
            ]{hyperref}
\usepackage{xcolor}
\usepackage{dcolumn}
\usepackage{bm}
\usepackage{bbding}
\usepackage{times}
\usepackage{datetime}
\usepackage{scrtime}
\usepackage{natbib}
\usepackage[shortlabels]{enumitem}
\usepackage{url}
\usepackage{makecell}
\usepackage{multirow}
\usepackage{booktabs}
\usepackage{ulem}
\usepackage{dynkin-diagrams}

\def\Re{\mathop{\textrm {Re}}}
\def\Im{\mathop{\textrm {Im}}}

\begin{document}

\title{Spontaneous polarized phase transitions and symmetry breaking in the imbalanced Dicke model}
\author{Jinling Lian}
\email{jllian@usx.edu.cn}
\affiliation{Department of Physics, Shaoxing University, Shaoxing 312000, China}
\affiliation{Zhejiang Engineering Research Center of MEMS, Shaoxing University, Shaoxing 312000, China}
\author{Ran Huang}
\affiliation{Department of Physics, Shaoxing University, Shaoxing 312000, China}
\author{Chao Gao}
\email{gaochao@zjnu.edu.cn}
\affiliation{Department of Physics, Zhejiang Normal University, Jinhua 321004, China}
\author{Lixian Yu}
\affiliation{Department of Physics, Shaoxing University, Shaoxing 312000, China}
\author{Qi-Feng Liang}
\email{qfliang@usx.edu.cn}
\affiliation{Department of Physics, Shaoxing University, Shaoxing 312000, China}
\affiliation{Zhejiang Engineering Research Center of MEMS, Shaoxing University, Shaoxing 312000, China}
\author{Wu-Ming Liu}
\affiliation{Beijing National Laboratory for Condensed Matter Physics, Institute of Physics, Chinese Academy of Sciences, Beijing 100190, China}
\affiliation{School of Physical Sciences, University of Chinese Academy of Sciences, Beijing 100190, China}

\begin{abstract}
We investigate the ground-state properties and quantum phase transitions of an ensemble consisting of $N$ four-level atoms within an optical cavity coupled to the single cavity mode and external laser fields. The system is described by an extended imbalanced Dicke model, in which the co- and counterrotating coupling terms are allowed to have different coupling strengths. Some novel polarized phases characterized by the phase differences between the cavity field or the atomic spin excitation and the Raman laser are found analytically. Meanwhile, the full phase diagram and quantum phase transitions are also revealed. Two sign-dependent coexisting phases have been identified, where the normal phase persists as a metastable state alongside the superradiant ground state. Finally, the breaking or restoration of the intrinsic symmetry in this system is addressed. It is found that besides the continuous $U(1)$ and discrete $\mathbb{Z}_2$ symmetries, the system also exhibits two reflection symmetries $\sigma_v$s, a central symmetry $C_2$ in the abstract position-momentum representation, and a discrete reflection parity-time ($\mathcal{PT}$) symmetry, a parameter exchange symmetry $\mathcal{T}_\mathrm{ex}$ in the parameter space. These additional symmetries are governed by two Coxeter groups.
\end{abstract}

\maketitle

\section{Introduction}

The quantum phase transitions (QPTs) of many-body systems with few degrees of freedom have paved a fundamental way to explore the features of quantum criticality and quantum correlations \cite{sachdev2011}. It is now at the center of research in many realms, such as quantum simulation \cite{Georgescu2014Mar}, quantum information theory \cite{Zanardi2008Oct}, quantum precision measurements \cite{Zhang2016Jun, Safronova2018Jun}, and quantum thermodynamics \cite{Campaioli2024Jul, Ferraro2018Mar}. As a prototypical many-body model of cavity quantum electrodynamics (QED), the Dicke model \cite{Dicke1954Jan}, which describes a large ensemble of identical two-level atoms coupled to a quantized cavity field, has attracted much attention. As is well-known, when the number of atoms tends to infinity, the model undergoes a QPT from the normal phase (NP) to the superradiant phase (SP), which belongs to the mean-field Ising universality class \cite{Kirton2019Feb} and undergoes a structural change \cite{Dogra2019Dec} in the properties of the ground-state energy spectrum at the critical threshold of the coupling between the atoms and the bosonic field \cite{Gilmore1978May, Emary2003a, Emary2003}. This type of phase transition is known as the Dicke superradiant phase transition and has been intensely discussed theoretically and experimentally \cite{Hepp1973, Gilmore1978May, Emary2003a, Emary2003, Dimer2007, Bastidas2012Jan, Baumann2010Apr, Baumann2011, Liu2011Mar, ZhangXF2013Feb, Baden2014, Zhiqiang2017Apr, Mlynek2014Nov, Safavi-Naini2018Jul, Bakemeier2012Apr, Zhu2016Jan, Gomez-Ruiz2016Sep, Kirton2019Feb, Lian2012Jun, WangZM2016Mar, Zhu2024May, Li2018Aug, Makihara2021May, Kim2025Apr}.

In addition, due to its apparent advantages, ultracold atomic ensemble coupled to a high-finesse optical cavity has become a versatile tool for investigating many-body physics \cite{Ritsch2013, Cooper2019Mar}. By mapping the self-organized behavior \cite{Domokos2002Dec, Nagy2008Jun, Black2003Nov} of a Bose-Einstein condensate coupled to an optical cavity into the Dicke model, the Dicke superradiant phase transition has been observed experimentally \cite{Baumann2010Apr, Baumann2011}. Also, subsequent theoretical studies predicted a variety of novel quantum phases as well as emergent ground-state and dynamical properties of the Dicke model on various scenarios \cite{Keeling2010Jul, Bhaseen2012Jan, Liu2011Mar, Zhang2013Jan, Grimsmo2013Nov, Kulkarni2013Nov, Oztop2012Aug, Nagy2011Oct, Buijsman2017Feb, MarquezPeraca2024Mar}. In particular, to access and expand the adjustable range and the independence of the parameter regimes, an effective imbalanced Dicke model is realized by cavity QED systems based on ultracold atoms through two cavity-assisted Raman transitions \cite{Baden2014, Zhiqiang2017Apr}. The most immediate consequence of the expansion and independence of the parameter regimes is the emergence of many novel phases, accompanying the rich phase diagram and symmetries \cite{YuYX2013Dec, Fan2014Feb, Baksic2014, Baden2014, Yu2015, Yu2016, Buijsman2017Feb, Zhiqiang2017Apr, Leonard2017Mar, Liu2017, Moodie2018Jan, Stitely2020Jul, Ferri2021Dec, Yu2022Aug, Ray2022Apr, Huang2023Jun, Chiacchio2023Sep, Zhu2024May, Soriente2018May, Soriente2020Feb, Soriente2021May, FallasPadilla2023Sep}. Especially, polarized phases \cite{Morales2019Jul} characterized by the phase differences in Bose gas \cite{Kroeze2018Oct, Guo2019MayPRL, Li2021Mar, Esslinger2022Aug} or Fermi gas \cite{Nie2023Oct} loaded in a single-mode cavity are identified. In these ultracold atomic gases, the phase difference plays a pivotal role, which is closely related to the self-organization behavior and the QPTs. Furthermore, QPTs in these specific models can also be characterized by Landau-type order parameters and are accompanied by the breaking of the intrinsic symmetry, such as the breaking of continuous $U(1)$ and discrete $\mathbb{Z}_2$ \cite{YuYX2013Dec, Fan2014Feb, Baksic2014, Yu2015, Yu2016, Leonard2017Mar, Liu2017, Moodie2018Jan, Stitely2020Jul}, as well as the $\mathcal{PT}$ symmetry \cite{Chiacchio2023Sep}.

In the present work, we primarily focus on the ground-state properties and QPTs of an ultracold atomic ensemble in an optical cavity coupled to the cavity mode and external laser fields by mean-field methods. Our main findings are presented below. (i) Through the scaled ground-state energy and the ground-state stability, we find a series of novel spontaneous polarized phases, which we dub $x$-SP/RSP, $p$-SP/RSP, SP$_0$/RSP$_0$, and SP$_{x}$/SP$_{p}$. These polarized phases are primarily characterized by the distinct behaviors of the phase difference. (ii) The order parameters, including the mean photon number, the scaled atom population, and the scaled dipole moments in different cases, are derived analytically. (iii) The critical lines of the QPTs among the NP, the SPs mentioned above, and two sign-dependent coexisting phases ($p$-SP/RSP+NP) are obtained analytically. (iv) Together with the rich novel phases and superradiant phase transitions, a full symmetry diagram is revealed. The breaking, partial breaking, or restoration of symmetries, involving well-known continuous $U(1)$ and discrete $\mathbb{Z}_2$, as well as crucial reflection symmetry $\sigma_v$ and central symmetry $C_2$ associated with a Coxeter group $W$, along with reflection parity-time symmetry $\mathcal{PT}$ and parameter exchange symmetry $\mathcal{T}_\mathrm{ex}$ associated with another Coxeter group $W^\prime$, are obtained.

The paper is organized as follows. The model and the effective Hamiltonian are introduced in Sec. \ref{Sec_Model}. In Sec. \ref{Sec_GS}, the ground-state properties and stability are obtained. In Sec. \ref{Sec_PolarizedPhases}, the polarized phases and QPTs in this particular model are revealed in different cases. The breaking of symmetries are discussed in detail in Sec. \ref{Sec_Symmetry}. Finally, the conclusions are given in Sec. \ref{Sec_Conclusion}.


\section{The model: ultracold atomic ensemble in the Raman-assisted cavity}
\label{Sec_Model}

\begin{figure}[b]
  \centering \vspace{0cm} \hspace{0cm} \scalebox{0.7}{
  \includegraphics{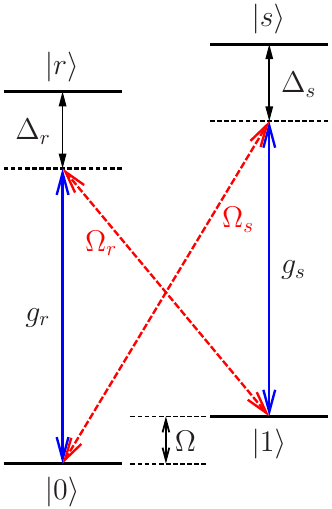}}
  \caption{(Color online) Energy levels of the atomic ensemble in a Raman-assisted cavity. The system consists of $N$ four-level atoms in a high finesse optical cavity coupled to the single cavity mode, where $|0\rangle$ and $|1\rangle$ are two ground-state sublevels of level-space $\Omega$. Two laser beams with Rabi frequencies $\Omega_r$ and $\Omega_s$, indicated by the red dashed arrows, are far detuned from the excited states but near-resonant with the cavity-assisted Raman transition, indicated by the blue solid arrows. In addition, the two excited states $|r\rangle$ and $ |s\rangle$ can be the same level.}
  \label{fig1}
\end{figure}

We start by considering an ensemble of $N$ four-level atoms in an optical cavity coupled to a single cavity mode and external laser fields. Figure \ref{fig1} illustrates the energy levels of the system \cite{Dimer2007, Arnold2012, Baden2014, Zhiqiang2017Apr}, in which $N$ rubidium-87 atoms are placed inside the mode volume of a high-finesse optical cavity and couple the $|F=1,m_{F}=1\rangle \equiv |0\rangle $ and $|F=2,m_{F}=2\rangle \equiv |1\rangle $ hyperfine ground states of level-space $\Omega$ via two cavity-assisted Raman transitions. Specifically, the Raman lasers couple the $|0\rangle \leftrightarrow |s\rangle$ and $|1\rangle \leftrightarrow |r\rangle$ transitions with Rabi frequencies $\Omega_r$ and $\Omega_s$, respectively, while the cavity mode couples the $|s\rangle \leftrightarrow |1\rangle$ and $|r\rangle \leftrightarrow |0\rangle$ transitions with strengths $g_r$ and $g_s$, respectively. The corresponding detunings are $\Delta_r $ and $\Delta_s$, as labeled in the figure.

After performing the adiabatic elimination of the excited states \cite{Dimer2007} and neglecting any off-resonant transitions \cite{Grimsmo2013}, the system can be modeled by the following Hamiltonian ($\hbar =1$) \cite{Dimer2007,Baden2014, Zhiqiang2017Apr}:
\begin{eqnarray}
H &=& \omega a^{\dagger }a+\left( \Omega +\frac{U}{N}a^{\dagger }a\right) J_{z}  \notag \\
&& +\frac{\lambda }{\sqrt{N}}\left( a^{\dagger }J_{-}+aJ_{+}\right) +\frac{\kappa }{\sqrt{N}}\left( a^{\dagger }J_{+}+aJ_{-}\right).
\label{H_system}
\end{eqnarray}
Here $a$ ($a^\dagger$) is the annihilation (creation) operator of the single radiation mode $\omega$. The collective spin operators are given by $J_+ = J_-^\dagger = \sum_{i=1}^N {b_{i,1}^\dagger b_{i,0}}$ and $J_z=\sum_{i=1}^N{(b_{i,1}^\dagger b_{i,1} - b_{i,0}^\dagger b_{i,0})/2}$, where $b_{i,0}$ ($b_{i,0}^\dagger$) and $b_{i,1}$ ($b_{i,1}^\dagger$) are the annihilation (creation) operators for the states $|0\rangle$ and $|1\rangle$ of the $i$th ultracold atom, respectively. $U$ is the nonlinear atom-field interaction. $\lambda \propto \sqrt{N} g_r\Omega_r/\Delta_r$ and $\kappa \propto \sqrt{N} g_s\Omega_s/\Delta_s$ denote the coupling strengths of the co- and counterrotating terms, respectively. It is important to note that this two-level description assumes large detunings from the excited states, such that the hyperfine structure remains unresolved, off-resonant Raman transitions are negligible, and motional excitations are adiabatically eliminated. These conditions are typically realized in deep optical lattice experiments \cite{Baden2014,Zhiqiang2017Apr}. In these cavity-assisted Raman coupling schemes, independent tuning of the co- and counterrotating terms is attained through the Rabi frequencies $\Omega_{r(s)}$ and detunings $\Delta_{r(s)}$. This enables the exploration of imbalanced Dicke models across a broad range of parameters, including sign reversal of the coupling strengths. Notably, a complementary mechanism for sign reversal has been reported in the spin-$1$ realization of the Dicke model when extended to the $F=2$ manifold, where the change in sign of the Clebsch–Gordan (CG) coefficients naturally flips the coupling parameters \cite{Zhiqiang2017Apr}.

The collective nature of the coupling between the atomic ensemble and the cavity field motivates the use of the Holstein-Primakoff (HP) transformation \cite{Holstein1940} which maps a collective spin-$N/2$ particle into an effective boson with associated creation and annihilation operators $c^\dagger$ and $c$, such that $J_{+}={J_{-}}^\dagger=c^{\dagger }C$ and $J_{z}=c^{\dagger }c-N/2$. Here the Hermitian operator $C=(N-c^{\dagger }c)^{1/2}$ allows the collective spin operators to satisfy the SU(2) algebra. Hamiltonian (\ref{H_system}) can be rewritten as
\begin{eqnarray}
H &=& \omega a^{\dagger }a+\left( \Omega +\frac{U}{N}a^{\dagger }a\right) \left( c^{\dagger }c-\frac{N}{2}\right) \notag \\
&&+\frac{\lambda }{\sqrt{N}}\left(a^{\dagger }C c+ac^{\dagger }C\right) +\frac{\kappa }{\sqrt{N}}\left( a^{\dagger }c^{\dagger }C+a C c\right).
\label{H_HP}
\end{eqnarray}
To deal with this model, we now introduce two \textit{shift-rotating} boson operators $\tilde{a} = a+\sqrt{N}\alpha $ and $\tilde{c} = c - \sqrt{N}\gamma$, where $\alpha = \rho \mathrm{e}^{\mathrm{i}\theta}$ and $\gamma = \mu \mathrm{e}^{\mathrm{i}\eta}$ with $\rho,\mu \geqslant 0$ and $\theta,\eta \in [0,2\pi)$ are complex auxiliary parameters to be determined and describe the collective excitations of the cavity mode and the ultracold atoms, respectively \cite{Emary2003a, Emary2003, Lian2013}. Furthermore, this assumption can be treated via the Wigner function of the boson coherent states $|\alpha\rangle$ and $|\gamma\rangle$ \cite{Lian2013}. For a mesoscopic atomic ensemble within the cavity, we substitute the shift-rotating boson operators into the Hermitian operator $C$ and expand in powers of $N^{-1/2}$, obtaining $C = \sqrt{N}\mathcal{C} - (\gamma\tilde{c}^\dagger + \gamma^*\tilde{c})/ (2\mathcal{C}) + O(N^{-1/2})$ with $\mathcal{C} = (1-\gamma^*\gamma)^{1/2}$ a dimensionless parameter. Higher-order terms in the expansion of $C$ would contribute only to the quantum fluctuations beyond the mean-field level and do not affect the ground-state energy in the thermodynamic limit $N\to\infty$ \cite{Emary2003a, Emary2003, Lian2013}.

Substituting the truncated form of $C$ in Eq. (\ref{H_HP}) and collecting terms by powers of $N^{1/2}$, the Hamiltonian of the system takes the form \cite{Klein1991, Frasca2004Sep}
\begin{equation}\label{H_Expand}
  H = N H_0 + N^{1/2} H_1 + O(N^0),
\end{equation}
where $H_{0}\!=\!\omega \alpha ^{\ast }\alpha \!+\! ( \Omega \!+\!U\alpha ^{\ast }\alpha ) ( \gamma ^{\ast }\gamma \!-\! 1/2 ) \!-\!2\mathcal{C} \zeta(\alpha,\alpha^\ast,\gamma,\gamma^\ast) $ and $H_{1} = [ -\omega \alpha -U\alpha ( \gamma ^{\ast }\gamma -1/2) +\lambda\gamma \mathcal{C} +\kappa\gamma ^{\ast } \mathcal{C}] \tilde{a}^{\dagger } +( \Omega \gamma +U\alpha ^{\ast }\alpha \gamma -\lambda\alpha \mathcal{C} -\kappa\alpha ^{\ast } \mathcal{C}+\gamma \zeta /\mathcal{C}) \tilde{c}^{\dagger }+\text{H.c.}$ with the function $\zeta(\alpha,\alpha^\ast,\gamma,\gamma^\ast) =\lambda \Re (\alpha ^{\ast }\gamma) +\kappa \Re (\alpha \gamma) $. It should be noted that the coefficients multiplying the operators $\tilde{a}^\dagger$, $\tilde{a}$, $\tilde{c}^\dagger$, and $\tilde{c}$ in $H_1$ are precisely the derivatives of $H_0$ with respect to the complex parameters; hence $H_1$ vanishes once the equilibrium conditions are imposed. The leading-order term $H_0$ therefore reflects the ground-state energy, while $H_2$ (of order $O(N^0)$), not shown here, governs the excitation spectrum \cite{Emary2003a, Emary2003}.

\section{Ground-state properties and stability}
\label{Sec_GS}
Physically, $H_0$ determines the scaled ground-state energy in terms of $\alpha$, $\alpha^\ast$, $\gamma$ and $\gamma^\ast$: $\bar{E}(\alpha,\alpha^\ast,\gamma,\gamma^\ast)=H_0(\alpha,\alpha^{\ast},\gamma,\gamma^\ast)$ \cite{Klein1991}; i.e.,
\begin{equation}\label{E1}
\bar{E} = \omega \alpha ^{\ast }\alpha +\left( \Omega +U\alpha ^{\ast }\alpha\right) \left( \gamma ^{\ast }\gamma - 1/2 \right) -2\mathcal{C} \zeta.
\end{equation}
In principle, nonzero values of $\alpha$ and $\gamma$, which minimize the scaled energy $\bar{E}$, suggest nonzero coherences of the bosonic field and the spontaneous polarization of the collective spin of the ground state. Those coherences play the roles of order parameters of the Dicke-type superradiant phase transition \cite{Baksic2014}. All of them, including the mean photon number $\langle a^{\dagger}a\rangle / N = \alpha^{\ast}\alpha $, the scaled atom population $\langle J_z \rangle /N = \gamma^{\ast}\gamma - \frac{1}{2} $, and the scaled atom dipole moments $\langle J_x \rangle /N = \frac{\mathcal{C}}{2} (\gamma^{\ast} +\gamma) $ and $\langle J_y \rangle /N = \frac{\mathcal{C} }{2\mathrm{i}} (\gamma^{\ast} - \gamma) $ \cite{ Castanos2009Jun} are relevant to the complex auxiliary parameters $\alpha$ and $\gamma$. Considering the parameters $\rho $, $\mu \geqslant 0$ and $\theta $, $\eta \in \lbrack 0,2\pi )$, we have $\langle a^{\dagger}a\rangle / N = \rho^2$, $\langle J_z \rangle /N =\mu^2-\frac{1}{2}$, $\langle J_x \rangle /N = \mathcal{C} \mu \cos\eta$, and $\langle J_y \rangle /N = -\mathcal{C} \mu \sin\eta$ with $\mathcal{C} = (1-\mu^2)^{1/2}$. The scaled ground-state energy (\ref{E1}) can be rewritten as
\begin{eqnarray}\label{E2}
\bar{E} = \omega \rho ^{2} + \left( \Omega + U\rho ^{2}\right) \left( \mu ^{2} - 1/2 \right) - 2\rho \mu \mathcal{C} \zeta_+
\end{eqnarray}
with the function 
\begin{equation}
  \zeta_{\pm }(\theta,\eta)=\left\{ 
  \begin{array}{ll}
    \lambda \cos ( \theta -\eta ) \pm \kappa \cos( \theta +\eta ) , & \rho\mu > 0 ,\\ 
    \lambda \pm \kappa, & \rho\mu=0.%
  \end{array}%
  \right. 
\end{equation}
Here, we introduce the auxiliary function $\zeta_-$ to simplify subsequent expressions. In the NP, the phase differences $\theta$ and $\eta$ are completely uncertain \cite{Baumann2011}. To derive the stability conditions, we thus set $\theta=\eta=0$ in the case of $\rho=\mu=0$ without loss of generality, yielding $\zeta_\pm = \lambda \pm \kappa$. The minimization of the ground-state energy (\ref{E2}) lies upon the equilibrium equations, which can be directly obtained by means of $\partial \bar{E} / \partial \circ = 0$ ($\circ \in \{\rho, \mu, \theta,\eta\} $); i.e.,
\begin{subequations}
\label{MidEqs}
\begin{align}
\omega \rho +U\rho \left( \mu ^{2}-1/2 \right) - \mu \mathcal{C} \zeta_{+} =0, \label{MidEqsA}\\
\left( \Omega +U\rho ^{2}\right) \mu - \rho (1-2\mu ^{2}) \mathcal{C}^{-1} \zeta_{+} =0, \label{MidEqsB}\\
\rho \mu \mathcal{C} \left[ \lambda \sin \left( \theta -\eta \right)+\kappa \sin \left( \theta +\eta \right) \right] =0, \label{MidEqsC}\\
\rho \mu \mathcal{C} \left[ \lambda \sin \left( \theta -\eta \right)-\kappa \sin \left( \theta +\eta \right) \right] =0. \label{MidEqsD}
\end{align}
\end{subequations}

Furthermore, the ground-state stability of the system is subject to a $4\times 4$ Hessian matrix $\mathcal{M}$, whose elements can be evaluated by the formula $\mathcal{M}_{ij}=\partial ^{2}E/( \partial \circ \partial \diamond) $ ($i,j=1,2,3,4$ and $\diamond \in \{\rho,\mu,\theta,\eta\}$). Explicitly, $\mathcal{M}$ is given by the expression
\begin{equation}
  \mathcal{M} = \mathrm{diag}[\mathcal{M}_A, \mathcal{M}_P],
  \label{Hessian}
\end{equation}
where $\mathcal{M}_A = [\mathcal{M}_{11},\mathcal{M}_{12};\mathcal{M}_{21},\mathcal{M}_{22}]$ and $\mathcal{M}_P = \mathcal{M}^\prime [\zeta_{+}, -\zeta_{-}; -\zeta_{-}, \zeta_{+} ]$ with $\mathcal{M}_{11} = 2\omega +2U(\mu ^{2}-1/2)$, $\mathcal{M}_{22} = 2\Omega +2U\rho^{2}+ 2\rho \mu (3-2\mu ^{2}) {\mathcal{C}^{-3}} \zeta_+$, $\mathcal{M}_{12} = \mathcal{M}_{21} = 4U\rho \mu -2(1-2\mu ^{2}) {\mathcal{C}^{-1}} \zeta_{+}$, and $\mathcal{M}^\prime = 2\rho \mu \mathcal{C} $. Obviously, Hessian matrix $\mathcal{M}$ consists of two real symmetric matrices, referred as $\mathcal{M}_A$ and $\mathcal{M}_P$, respectively. Due to the close correlation between $\mathcal{M}$ and the ground-state stability of the system, the total parameter space depending on $\lambda$ and $\kappa$ can then be divided into two equivalent subspaces, that are associated with the stability of the so-called \textit{amplitude-} and \textit{phase}-sectors, respectively.

Eigenvalues $m$ of the Hessian matrix (\ref{Hessian}) can be obtained analytically by
\begin{subequations}
\label{HM_m}
\begin{align}
m_{1}=& \frac{1}{2}\left( \mathcal{M}_{11} \! + \! \mathcal{M}_{22} \! - \! \sqrt{(\mathcal{M}_{11} \! - \! \mathcal{M}_{22})^{2} \! + \! 4 \mathcal{M}_{12}^{2}}\right),  \label{HM_m1} \\
m_{2}=& \frac{1}{2}\left( \mathcal{M}_{11} \! + \! \mathcal{M}_{22} \! + \! \sqrt{(\mathcal{M}_{11} \! - \! \mathcal{M}_{22})^{2} \! + \! 4 \mathcal{M}_{12}^{2}}\right),  \label{HM_m2} \\
m_{3}=& \mathcal{M}^\prime \left(\zeta_{+}+\zeta_{-}\right) = 2\lambda \mathcal{M}^\prime \cos\left( \theta -\eta \right),  \label{HM_m3} \\
m_{4}=& \mathcal{M}^\prime \left(\zeta_{+}-\zeta_{-}\right) = 2\kappa  \mathcal{M}^\prime \cos\left( \theta +\eta \right).  \label{HM_m4}
\end{align}
\end{subequations}
If $\mathcal{M}$ is positive definite (i.e., all eigenvalues $m$ of $\mathcal{M}$ are positive), the scaled ground-state energy $\bar{E}$ has local minimums and the system is located at the stable phases. If $\mathcal{M}$ is indefinite (i.e., some eigenvalues $m>0$, while some $m<0$), the scaled ground-state energy $\bar{E}$ has saddle points and the system is dynamically unstable. If $\mathcal{M}$ is negative definite (i.e., all eigenvalues $m$ of $\mathcal{M}$ are negative), the scaled ground-state energy $\bar{E}$ has a local maximum and the system is extremely unstable \cite{Lian2013}. Obviously, the phase boundaries of the stable phases can be obtained in terms of the condition $\min {m}=0$ ($\mathcal{M}$ is a positive semidefinite matrix by now).

\begin{table*}[t]
  \caption{Relations among the auxiliary function $\zeta_+$, phase differences $[\theta,\eta]$, and the order parameters in different phases of the system for nonzero $\kappa$. For brevity, we define $\mathcal{L}_x = {(1-\lambda^4_{c,x}/\lambda^4)}^{1/2}$ and $\mathcal{L}_p ={(1-\lambda^4_{c,p}/\lambda^4)}^{1/2}$.}
  \label{table_phase2}
  \setlength{\tabcolsep}{2.5mm}
  \resizebox{0.8\linewidth}{!}
  {
  \begin{tabular}{l c c c c c c}
    \specialrule{0.08em}{3pt}{3pt}
    Phase & $\zeta_{+}$ & $\left[ \theta ,\eta \right] $ & $\left\langle a^{\dagger}a \right\rangle /N $ & $\left\langle J_{z}\right\rangle /N$ & $\left\langle J_{x}\right\rangle /N$ & $\left\langle J_{y}\right\rangle /N$ \\ \specialrule{0.05em}{3pt}{5pt}

    NP & $-$ & $-$ & $0$ & $-\frac{1}{2}$ & $0$ & $0$ \\  \specialrule{0em}{2pt}{2pt}
    
    $x$-SP & $\lambda \! + \! \kappa $ & $\left[ 0,0\right] ,\left[ \pi ,\pi \right] $ & $\frac{\left( 1+t\right) ^{2}\lambda ^{2}}{4\omega ^{2}}\mathcal{L}^2_x $ & $-\frac{\lambda _{c,x}^{2}}{2\lambda ^{2}}$ & $ \frac{1}{2}\mathcal{L}_x, - \frac{1}{2}\mathcal{L}_x$ & $0,0$ \\  \specialrule{0em}{2pt}{2pt}
    $x$-RSP & $- \! \left(  \! \lambda  \! + \! \kappa \!  \right) $ & $\left[ 0,\pi \right] ,\left[ \pi,0\right] $ & $\frac{\left( 1+t\right) ^{2}\lambda ^{2}}{4\omega ^{2}}\mathcal{L}^2_x $ & $-\frac{\lambda_{c,x}^{2}}{2\lambda ^{2}}$ & $ -\frac{1}{2}\mathcal{L}_x, \frac{1}{2}\mathcal{L}_x$ & $0,0$ \\  \specialrule{0em}{2pt}{2pt}
    $p$-SP & $\lambda  \! - \! \kappa $ & $\left[  \! \frac{\pi }{2} \! , \! \frac{\pi }{2} \! \right] , \left[  \! \frac{3\pi }{2} \! , \! \frac{3\pi }{2} \! \right] $ & $\frac{\left( 1-t\right)^{2}\lambda ^{2}}{4\omega ^{2}}\mathcal{L}^2_p $ & $-\frac{\lambda _{c,p}^{2}}{2\lambda ^{2}}$ & $0,0$ & $ -\frac{1}{2}\mathcal{L}_p, \frac{1}{2}\mathcal{L}_p$ \\  \specialrule{0em}{2pt}{2pt}
    $p$-RSP & $- \! \left( \!  \lambda  \! - \! \kappa  \! \right) $ & $\left[  \! \frac{\pi }{2} \! , \! \frac{3\pi}{2} \! \right] ,\left[  \! \frac{3\pi }{2} \! , \! \frac{\pi }{2} \! \right] $ & $\frac{\left(1-t\right) ^{2}\lambda ^{2}}{4\omega ^{2}}\mathcal{L}^2_p $ & $-\frac{\lambda _{c,p}^{2}}{2\lambda ^{2}}$ & $0,0$ & $ \frac{1}{2}\mathcal{L}_p, -\frac{1}{2}\mathcal{L}_p$ \\ \specialrule{0.08em}{3pt}{3pt}
  \end{tabular}
  }
\end{table*}

\begin{figure}[b]
  \centering \vspace{0cm} \hspace{0cm} \scalebox{0.55}{%
  \includegraphics{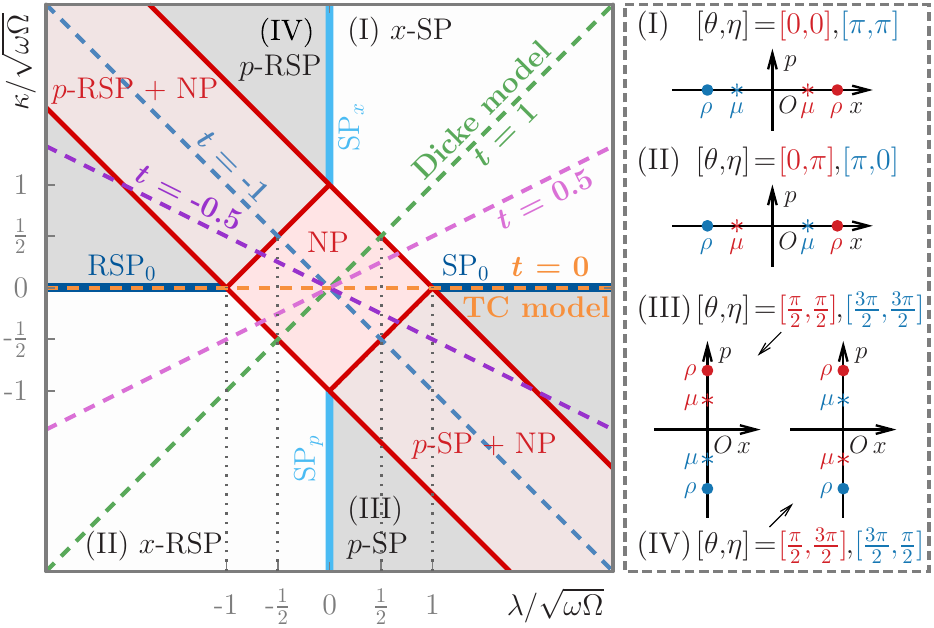}}
  \caption{(Color online) Phase diagram of the system versus the rescaled atom-photon interactions $\lambda/\sqrt{\omega\Omega}$ and $\kappa/\sqrt{\omega\Omega}$ (left panel). According to the configurations of $x_a(p_a)$ and $x_c(p_c)$, the SP is mainly divided into four regions, namely, (I) $x$-SP, (II) $x$-RSP, (III) $p$-SP, and (IV) $p$-RSP. The critical lines $\kappa_c = \pm(\sqrt{\omega\Omega} - \vert {\lambda_c}\vert)$, as plotted by the red solid lines, separate the NP from the SPs. Configurations of $x_a(p_a)$ and $x_c(p_c)$ corresponded with various phase differences $[\theta,\eta]$ in the position-momentum representation (right panel), which are marked by $\bullet$ and $\ast$, respectively. The dashed lines indicate five special circumstances ($t=0, \pm 1, \pm 0.5$), which will be discussed in detail below. In addition, the SPs on the dark and light blue solid lines, denoted by RSP$_0$/SP$_0$ and SP$_x$/SP$_p$ respectively, behave differently from the $x$-SP/RSP and $p$-SP/RSP.}
  \label{fig_phase}
\end{figure}

Before revealing the ground-state properties of the model described by Eq. (\ref{H_system}), let us first consider the relations between the auxiliary parameters $\theta$ and $\eta$, and the function $\zeta_+$ compatible with them. Physically, $\theta$ ($\eta$), which can be obtained experimentally \cite{Baumann2011, Li2021Mar, Esslinger2022Aug, Guo2019MayPRL, Guo2019MayPRA}, is indeed the phase difference between the Raman laser and the cavity field (the atomic spin excitation) \cite{Chen2010, Baumann2011}. Moreover, just as the experiment shows, NP with zero $\rho$ is always accompanied by a completely uncertain $\theta$ \cite{Baumann2011}. So does the relationship between a zero $\mu$ and its rotating angle $\eta$. As a consequence, the Hessian matrix $\mathcal{M}$ has a rank of 2 in this circumstance. Therefore the stability of the system in NP totally depends upon the first two eigenvalues $m_{1,2}$, while the last two $m_{3,4}=0$ are eliminated physically. Meanwhile, the parameter space of NP would reduce into the \textit{amplitude}-sector immediately. On the other hand, for the stable SP with nonzero $\rho$ and $\mu$, it is concluded that $\sin(\theta \pm \eta ) =0$ according to Eqs. (\ref{MidEqsC}) and (\ref{MidEqsD}). Thus we have $\theta ,\eta \in \{0,\pi\}$ or $\{\pi /2, 3\pi/2\}$. As a direct result, values of the function $\zeta_{+}(\theta,\eta)$ in (\ref{MidEqsA}) and (\ref{MidEqsB}), which rely upon the terms $\cos( \theta \pm \eta ) $, are completely determined by means of the combinations of $\theta$ and $\eta$. Specifically, $\zeta_+$ with different $\theta $ and $\eta $ can be obtained as follows: (I) $\zeta_{+}=\lambda +\kappa $ with $[\theta ,\eta ]=[0,0]$ or $[\pi ,\pi ]$; (II) $\zeta_{+}=-(\lambda + \kappa) $ with $[\theta ,\eta ]=[0,\pi ]$ or $[\pi,0] $; (III) $\zeta_{+}=\lambda -\kappa $ with $[\theta ,\eta ]=[\pi /2,\pi /2]$ or $[3\pi /2,3\pi /2]$; or (IV) $\zeta_{+}=-(\lambda - \kappa) $ with $[\theta ,\eta ]=[\pi /2,3\pi/2]$ or $[3\pi /2,\pi /2]$ (as shown in Table \ref{table_phase2} and the right panel of Fig. \ref{fig_phase}). The discreteness of $\theta$ and $\eta$ should be attributed to the anisotropy of the atomic self-organization, which is induced by the imbalanced coupling terms of the atom-cavity system \cite{Guo2019MayPRL, Li2021Mar, Esslinger2022Aug}. It should be pointed out that, the aforementioned conclusions about the values of $\theta$ and $\eta$ are drawn under a nonzero $\kappa$. For zero $\kappa$, it is evident from Eqs. (\ref{MidEqsC}) and (\ref{MidEqsD}) that $\sin(\theta-\eta)=0$ for the stable SP. These equations may thus be fulfilled by taking $\theta = \eta$ or $\eta \pm \pi$ and the corresponding $\zeta_+$ then would be $\lambda$ or $-\lambda$ respectively. We will discuss this situation in more detail below. Obviously, different $\zeta_+$ arising from the different combinations of the phase differences $\theta$ and $\eta$ will directly influence the ground-state properties of the system.

To further clarify the role of the phase differences, we adopt an abstract position-momentum representation for each of the boson operators $a$($\tilde{a}$) and $c$($\tilde{c}$) via $x_a \equiv(1/\sqrt{2\omega})(a^\dagger + a)$, $p_a \equiv \mathrm{i} \sqrt{\omega/2}(a^\dagger - a)$, $x_c \equiv(1/\sqrt{2\Omega})(c^\dagger + c)$, and $p_c \equiv \mathrm{i} \sqrt{\Omega/2}(c^\dagger - c)$ \cite{Emary2003a, Emary2003}. Considering the fact that the action of the parity-like operator $\Pi = \exp [\mathrm{i}(\theta a^\dagger a + \eta c^\dagger c)]$ corresponds to rotations by the proper $\theta$ and $\eta$ about the origin in this representation, we obtain $x_a \sim \sqrt{N}\Re{\alpha} = \sqrt{N} \rho\cos\theta$, $p_a \sim \sqrt{N}\Im{\alpha} = \sqrt{N} \rho \sin\theta$, $x_c \sim \sqrt{N}\mu \cos\eta$, and $p_c \sim \sqrt{N} \mu \sin\eta$ in their respective quadrants of the $x$-$p$ representation. A similar procedure within the Wigner function also works \cite{Lian2013}. Obviously, when the system is in NP, $x_a(p_a)$ and $x_c(p_c)$ must be zero and the corresponding phase differences $\theta$ and $\eta$ are totally uncertain in the meantime. Besides this trivial situation, a detailed discussion of the characteristics of $x_a(p_a)$, $x_c(p_c)$ and $\theta$, $\eta$ in SP is required. As shown in Fig. \ref{fig_phase}, for the cases (I) and (II) with $\zeta_+ = \pm (\lambda + \kappa)$, $\theta, \eta \in \{0, \pi\}$ and the undetermined parameters $\alpha$, $\gamma$ are real, which implies the system may undergo a $x$-type superradiant phase ($x$-SP); while for the cases (III) and (IV) with $\zeta_+ = \pm (\lambda - \kappa)$, $\theta, \eta \in \{\pi/2, 3\pi /2 \}$ and the parameters $\alpha$, $\gamma$ are imaginary numbers, then the system may undergo a $p$-type superradiant phase ($p$-SP) \cite{Liu2017}. As illustrated in the right panel of Fig. \ref{fig_phase}, it is found that $\theta = \eta$ in (I) and (III); while $\theta = \eta \pm \pi$ in (II) and (IV). These particular results are also listed in Table \ref{table_phase2}. To characterize the \textit{reversal} behavior of $\theta$ and $\eta$ in case II (IV), which directly leads to the result that $x_a x_c<0$ ($p_a p_c <0$), the phase of case II (IV) is referred to as $x$-RSP ($p$-RSP). Note that the phase differences $\theta$ and $\eta$ take only discrete values from $\{0,\pi/2,\pi,3\pi/2\}$, so the cavity field and the atomic spin excitation are totally polarized, i.e., all SPs are polarized phases and the system may undergo a polarized QPT along with the change of the parameters. It is worth noting that, although the magnitudes of the order parameters $|\langle J_x\rangle/N|$ and $|\langle J_y\rangle/N|$ are identical between $x$-SP and $x$-RSP (as well as between $p$-SP and $p$-RSP), these pairs represent physically distinct phases distinguished by their phase-difference configurations $[\theta,\eta]$: $x$-SP ($p$-SP) exhibits in-phase locking ($\theta=\eta$), while $x$-RSP ($p$-RSP) exhibits out-of-phase locking ($\theta-\eta=\pm\pi$). This distinction manifests as opposite signs of $x_a x_c$ (or $p_a p_c$ for the $p$-type phases) in the position-momentum representation (see the right panel of Fig. \ref{fig_phase}).

Furthermore, according to the conditions of the ground-state stability, Eqs. (\ref{HM_m3}) and (\ref{HM_m4}) require $\lambda\cos(\theta-\eta) > 0$ and $\kappa\cos(\theta+\eta)>0 $. As a result, coupling parameters $\lambda$ and $\kappa$ should be both positive in $x$-SP, while both negative in $x$-RSP; and $\lambda > 0$ and $\kappa<0$ in $p$-SP, while $\lambda<0$ and $\kappa>0$ in $p$-RSP (i.e., $\zeta_+>0$ is always satisfied in SPs), as shown in Fig. \ref{fig_phase}. The stability of these potential SPs within the \textit{phase}-sector of the parameter space is directly dependent on the signs of $\lambda$ and $\kappa$. Consequently, the discrete phase-locking configurations are controlled by the signs of $\lambda$ and $\kappa$, which can be experimentally tuned via the detunings $\Delta_{r(s)}$, or alternatively, preset by choosing a hyperfine manifold with opposite CG coefficients~\cite{Baden2014, Zhiqiang2017Apr}. Thus, the $x$-SP/$x$-RSP pair (and analogously the $p$-SP/$p$-RSP pair) shares the same amplitude structure but is distinguished by whether the cavity field and the atomic spin excitation are in phase ($|\theta-\eta|=0$) or out of phase ($|\theta-\eta|=\pi$). Similar findings have been experimentally observed in several self-organizing atom-cavity systems \cite{Baumann2011, Li2021Mar, Esslinger2022Aug, Guo2019MayPRL}.

\section{Polarized phases and quantum phase transitions}
\label{Sec_PolarizedPhases}

\begin{figure}[b]
  \centering\includegraphics[width = 1.0\linewidth]{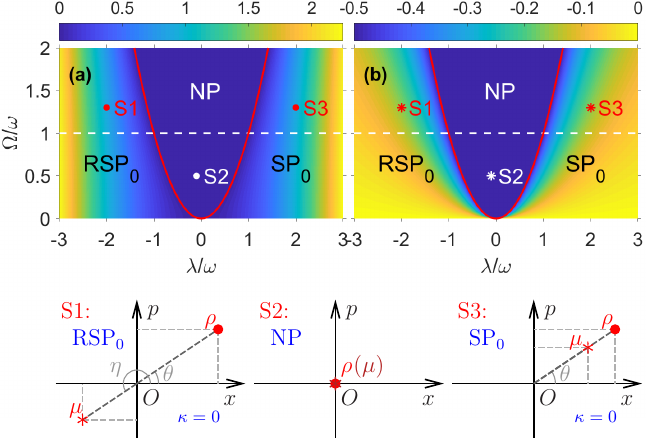}
  \caption{(Color online) (a) The mean photon number $\langle a^\dagger a \rangle /N$ and (b) the scaled atom population $\langle J_z \rangle /N$ as functions of the collective atom-photon coupling strength $\lambda$ and the atom frequency $\Omega$ with respect to $t=0$ ($\kappa=0$), respectively. The bottom panels S1-S3 reflect the complex amplitudes in the abstract position-momentum representation. For the TC model in this case, the changing of the phase differences $\theta$, $\eta \in [0, 2\pi)$ can be continuous in SP$_0$ and RSP$_0$, which is very different from the discrete ones in the cases of nonzero $\kappa$ (as shown in Table \ref{table_phase2}).}
  \label{fig2}
  \end{figure}

According to Eqs. (\ref{MidEqs}) and (\ref{HM_m}), the nonlinear atom-photon interaction $U$ only comes into play in the \textit{amplitude}-sector. It does not affect the phase differences but modifies the phase boundary and introduces an unstable region into the phase diagram. To characterize the spontaneous polarized phases of the system, it is convenient to neglect the nonlinear atom-photon interaction ($U=0$). Hamiltonian (\ref{H_system}) becomes
\begin{equation}
H=\omega a^{\dagger }a+\Omega J_{z}+\frac{\lambda }{\sqrt{N}} \! \left(
a^{\dagger }J_{-} \!+\! aJ_{+}\right) +\frac{\kappa }{\sqrt{N}} \! \left( a^{\dagger
}J_{+} \!+\! aJ_{-}\right),  \label{H_0}
\end{equation}%
which models the interaction of a four-level atomic ensemble pumped by a couple of Raman lasers with a cavity field via dipole interactions within an ideal cavity. The equilibrium equations (\ref{MidEqsA}) and (\ref{MidEqsB}) then become $\rho = \zeta_{+}\mu \sqrt{1-\mu ^{2}}/\omega $ and $\mu ( S_0 \mu ^{2}+T_0 )=0$ with $S_0= 2\omega {\zeta_{+}^{2}} $ and $T_0 =  \omega^2 \Omega  - \omega \zeta_{+}^{2} $. It is easy to find $\mu ^{2}=0$ or $\frac{1}{2} ( 1-{\omega \Omega }/{\zeta_{+}^{2}})$ for $\omega \Omega \leqslant \zeta_{+}^{2}$.

\begin{figure}[t]
  \centering\includegraphics[width = 1.0\linewidth]{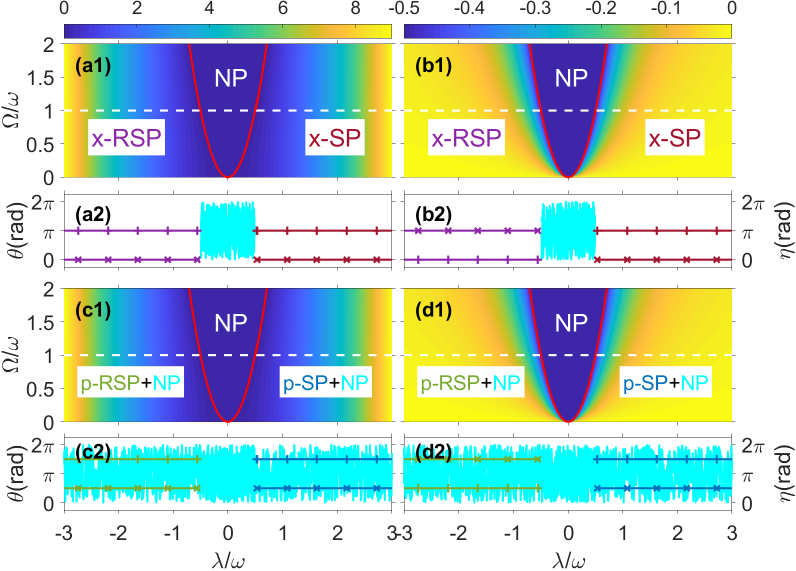}
  \caption{(Color online) (a1), (c1) The mean photon number $\langle a^\dagger a \rangle /N$ and (b1), (d1) the scaled atom population $\langle J_z \rangle /N$ as functions of the collective atom-photon coupling strength $\lambda$ and the atom frequency $\Omega$ with respect to $t=1$ (upper panels (a1) and (b1)) and $t=-1$ (lower panels (c1) and (d1)), respectively. (a2-d2) reflect the changing of the corresponding phase differences $\theta$ and $\eta$ as functions of $\lambda$ along a selected line $\Omega=\omega$ (the white dashed lines).}
  \label{fig3}
\end{figure}

For simplicity, we suppose $\kappa = t\lambda$ ($t$ is real) and consider the different situations of the ratio $t$:

(i) When $t=0$, Hamiltonian (\ref{H_0}) becomes $H=\omega a^{\dagger }a+\Omega J_{z}+(\lambda / \sqrt{N}) ( a^{\dagger }J_{-}+aJ_{+} ) $, which is exactly the so-called Tavis-Cummings (TC) model \cite{Tavis1968} for a weak-coupling system under rotating wave approximation (RWA) (see the orange dashed line in Fig. \ref{fig_phase}). It is obtained $[ \rho ,\mu^{2}] =[0,0] $ (NP) or $[\frac{|\lambda|}{2\omega } \sqrt{1-\lambda _{c}^{4}/\lambda ^{4}},\frac{1}{2} ( 1-\lambda_{c}^{2}/\lambda ^{2}) ]$ (SP) with $\left\vert \lambda_{c}\right\vert =\sqrt{\omega \Omega} $ being the critical points. Also, owing to $\kappa=0$, we have $\theta = \eta$ or $\eta \pm \pi$, and $\zeta_+ = \lambda\cos(\theta-\eta) =| \lambda|$. Hence ($x_a$, $x_c$, $p_a$, $p_c$)$\sim$($\rho\cos\theta$, $\pm \mu\cos\theta$, $\rho\sin\theta$, $\pm \mu\sin\theta$). As the bottom panels S1-S3 of Fig. \ref{fig2} shown, the phase difference $\theta$ ($\eta$) in SP is an arbitrary angle in $[0,2\pi)$ and the complex amplitudes of the boson operators $a$($\tilde{a}$) and $c$($\tilde{c}$) deviate from the origin along the opposite or same direction of $\theta$. That is to say, as the system evolves along the orange dashed lines in Fig. \ref{fig_phase}, the stable SP would exhibit some different behavior. Similarly, as indicated by the dark blue solid lines in Fig. \ref{fig_phase}, the corresponding phases are named after RSP$_0$ or SP$_0$, respectively. Fig. \ref{fig2} (a) and (b) provide the mean photon number $\langle a^\dagger a \rangle /N$ and the scaled atom population $\langle J_z \rangle /N$ as functions of the collective atom-photon coupling strength $\lambda$ and the atom frequency $\Omega$ with respect to $t=0$, respectively. According to the figure, as $\lambda$ increases, the system undergoes QPTs from RSP$_0$ to NP to SP$_0$ continuously. The red lines ($|\lambda_c|=\sqrt{\omega \Omega}$) indicate the phase boundary between NP and SP. Moreover, the white dashed lines reflect the evolution of the system in resonance situation ($\Omega = \omega$). 

(ii) When $t=1$, Hamiltonian (\ref{H_0}) becomes $H=\omega a^{\dagger }a+\Omega J_{z}+(\lambda /\sqrt{N})( a^{\dagger }+a) ( J_{+}+J_{-}) $, which is precisely the standard Dicke Hamiltonian (DH). Due to $\zeta_+ = |2\lambda| $, we could directly obtain $[\rho,\mu ^{2}] =[ 0,0]$ (NP) or $[\frac{|\lambda|}{\omega}\sqrt{1-\lambda_{c}^{4}/\lambda^{4}},\frac{1}{2}(1-\lambda _{c}^{2}/\lambda ^{2}) ]$ and $\theta, \eta \in \{0,\pi\}$ ($x$-SP/RSP) with $\left\vert \lambda_{c}\right\vert =\sqrt{\omega \Omega }/2$ being the critical points (see the green dashed line in Fig. \ref{fig_phase}). Fig. \ref{fig3} (a1) and (b1) display $\langle a^\dagger a \rangle /N$ and $\langle J_z \rangle /N$ as functions of $\lambda$ and $\Omega$ with respect to $t=1$. It is shown that both the photon number and atom population undergo a transition of $x$-RSP$\rightarrow$NP$\rightarrow$$x$-SP. The results are identical to those in the Dicke model, except for more details. Meanwhile, Fig. \ref{fig3} (a2) and (b2) reveal the changing of the phase differences $\theta$ and $\eta$ along with the QPTs. Obviously, $\theta$ and $\eta$ are disordered in NP; and both of them are $0$ or $\pi$ simultaneously in $x$-SP \cite{Baumann2011}. However, the status will reverse completely in $x$-RSP. For $\theta=0$ or $\pi$, $\eta$ would be $\pi$ or $0$ respectively, as shown in Table \ref{table_phase2}.

(iii) When $t=-1$, Hamiltonian (\ref{H_0}) becomes $H=\omega a^{\dagger }a+\Omega J_{z}+\lambda /\sqrt{N} ( a^{\dagger }-a) ( J_{-}-J_{+}) $, which we dub the anti-Dicke model, behaves very similarly to DH (see the blue dashed line in Fig. \ref{fig_phase}). Concretely, due to $\zeta_+ = \pm(\lambda-\kappa) = \pm 2\lambda$ and $\theta, \eta \in \{\pi/2,3\pi/2\}$, the system exhibits very similar ground-state behavior as in the case of $\theta, \eta \in \{0,\pi \}$ of $t=1$ and is likely to undergo a transition from NP to $p$-SP/RSP with the critical points $|\lambda_c| = \sqrt{\omega\Omega}/2$. This intriguing phenomenon should be attributed to the presence of the phase differences $\theta $ and $\eta $, which change the way of the symmetry breaking of the system. Differing from the previous case of $t=1$, the system undergoes a two-phase ($p$-RSP+NP or $p$-SP+NP) coexisting equilibrium evolution (see Fig. \ref{fig3} (c1) and(d1)). This consequence can be explained through the ground-state stability determined by the Hessian matrix $\mathcal{M}$. As mentioned before, the stability of NP totally depends on the first two eigenvalues $m_{1,2}$ of $\mathcal{M}$. As a result of $\rho=\mu=0$, it is found that $m_{1,2} = \omega + \Omega \mp \sqrt{(\omega-\Omega)^2+4\lambda^2(1+t)^2}$. If $t=1$, it is required that the minimum of the eigenvalues $\min{m_{1,2}} = \omega + \Omega - \sqrt{(\omega-\Omega)^2+16\lambda^2} \geqslant 0$, which demands $|\lambda| \leqslant \sqrt{\omega\Omega}/2  $; while if $t=-1$, $\min{m_{1,2}} = \omega + \Omega - |\omega-\Omega| \geqslant 0$, which is always satisfied for non-negative $\omega$ and $\Omega$. So NP is always stable and the system undergo a coexisting phase in this case. Similar behavior of the phase differences $\theta$ and $\eta$ also occur in the coexistence regions, as shown in Fig. \ref{fig3} (c2) and (d2).

\begin{figure}[tp]
  \centering\includegraphics[width = 1.0\linewidth]{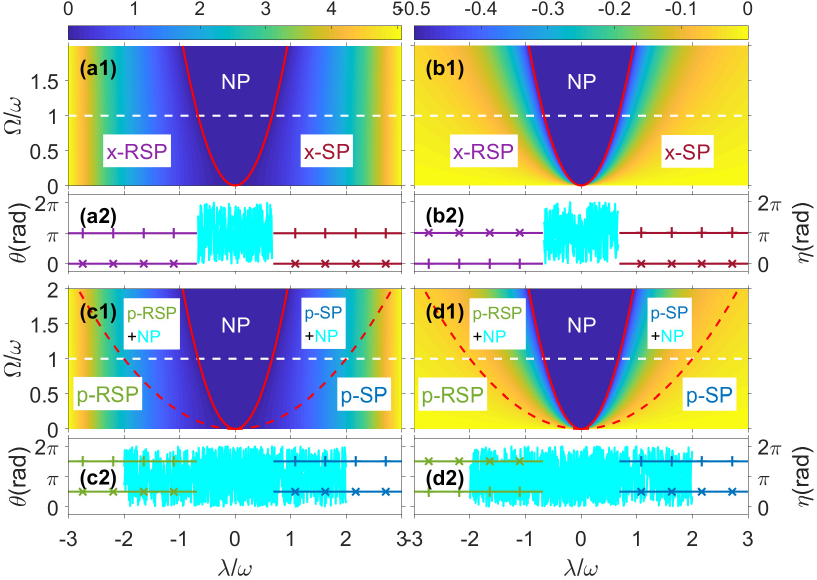}
  \caption{(Color online) (a1), (c1) The mean photon number $\langle a^\dagger a \rangle /N$ and (b1), (d1) the scaled atom population $\langle J_z \rangle /N$ as functions of the collective atom-photon coupling strength $\lambda$ and the atom frequency $\Omega$ with $t=0.5$ (upper panels (a1) and (b1)) and $t=-0.5$ (lower panels (c1) and (d1)), respectively. (a2-d2) reflect the corresponding phase differences $\theta$ and $\eta$ as functions of $\lambda$ along a selected line $\Omega = \omega$ (the white dashed lines). The red solid lines and the red dashed lines indicate the critical boundaries among different phases.}
  \label{fig_t_fraction}
\end{figure}

(iv) When $t \neq 0$ and $\pm 1$, in this imbalanced situation, it is found $ [ \rho ,\mu ^{2}] =[ 0,0] $ (NP) or $[\frac{|(1+t) \lambda |}{2\omega }\sqrt{1-\lambda _{c,x}^{4}/\lambda ^{4}}, \frac{1}{2}( 1-\lambda _{c,x}^{2}/\lambda ^{2}) ]$ ($x$-SP/RSP) with $\vert \lambda _{c,x}\vert =\sqrt{\omega \Omega }/\vert 1+t\vert $ being the critical points for $\theta, \eta \in \{0,\pi\}$. And, the counterparts in the other situation of $\theta, \eta \in \{\pi/2,3\pi/2\}$ can be obtained by means of a slight change: $[ \rho ,\mu ^{2}] =[ 0,0] $ (NP) or $[\frac{|( 1-t) \lambda |}{2\omega }\sqrt{1-\lambda_{c,p}^{4}/\lambda ^{4}},\frac{1}{2}( 1-\lambda _{c,p}^{2}/\lambda^{2}) ]$ ($p$-SP/RSP) with $\vert \lambda _{c,p}\vert =\sqrt{\omega \Omega}/\vert 1-t\vert $ being the critical points. In addition, the ground-state stability of the system can be obtained by means of the Hessian matrix (\ref{Hessian}) immediately. According to Eqs. (\ref{HM_m}), eigenvalues $m$ of $\mathcal{M}$ in this case can be reduced to $m_{1,2}= \omega +\Omega \mp \sqrt{\left( \omega -\Omega \right)^{2}+4\zeta_{+}^{2}}$ for NP. And
\begin{subequations}
\label{HM_m_U0_SP}
\begin{align}
m_{1,2}&= \omega +\frac{2\zeta_{+}^{4}}{\omega \left( \zeta_{+}^{2}+\omega \Omega \right) } \mp \sqrt{\mathcal{R}}, \label{HM_m_U0_SPA}\\
m_{3}&= \lambda \cos \left( \theta -\eta \right) \frac{\zeta_{+}}{\omega }\left(1-\frac{\omega ^{2}\Omega ^{2}}{\zeta_{+}^{4}}\right) , \label{HM_m_U0_SPB}\\
m_{4}&= \kappa  \cos \left( \theta +\eta \right) \frac{\zeta_{+}}{\omega }\left(1-\frac{\omega ^{2}\Omega ^{2}}{\zeta_{+}^{4}}\right)  \label{HM_m_U0_SPC}
\end{align}
\end{subequations}
with $\mathcal{R}=\omega ^{2}+(8\omega ^{2}\Omega ^{2}-4\zeta _{+}^{4})/(\omega \Omega +\zeta _{+}^{2})+4\zeta _{+}^{8}/[\omega ^{2}(\omega \Omega + \zeta _{+}^{2})^{2}]$. Furthermore, due to $\theta, \eta \in \{0,\pi\}$ in $x$-SP/RSP, Eqs. (\ref{HM_m_U0_SPB}) and (\ref{HM_m_U0_SPC}) reduced to $[m_{3}, m_{4}]= \mathcal{R}_x [\lambda, \kappa] $ with $\mathcal{R}_x = \frac{ \lambda(1+t)}{\omega}(1 - \lambda_{c,x}^4/\lambda^{4})$; while the counterparts in $p$-SP/RSP ($\theta, \eta \in \{\pi/2,3\pi/2\}$) reduce to $[m_{3},m_{4}] = \mathcal{R}_p [\lambda, -\kappa] $ with $\mathcal{R}_p = \frac{ \lambda (1-t)}{\omega} (1 - \lambda_{c,p}^4/\lambda^{4})$. They determine the boundary among NP, $x$-SP/RSP, $p$-SP/RSP, and the possible coexisting phase with (\ref{HM_m_U0_SPA}) together. Fig. \ref {fig_t_fraction} (a1) - (d1) illustrate $\langle a^\dagger a \rangle/N$ and $\langle J_z \rangle /N$ as functions of $\lambda$ and $\Omega$ with the ratios $t=\pm 0.5$. When $t = 0.5$, the system undergoes a QPT of $x$-RSP $\rightarrow$ NP $\rightarrow$ $x$-SP, just as in the case of $t=1$, except for a slight change in the position of the critical line; while the phase differences $\theta$ and $\eta$ also behave similarly. However, when $t=-0.5$, the system would undergo a transition of $p$-RSP $\rightarrow$ $p$-RSP+NP $\rightarrow$ NP $\rightarrow$ $p$-SP+NP $\rightarrow$ $p$-SP. The red solid lines and the red dashed lines in Fig. \ref{fig_t_fraction} (c1) and (d1) reveal the phase boundaries; while the phase differents $\theta$ and $\eta$ behave similarly and also show coexisting phenomena.

\begin{figure}[b]
  \centering\includegraphics[width = 1.0\linewidth]{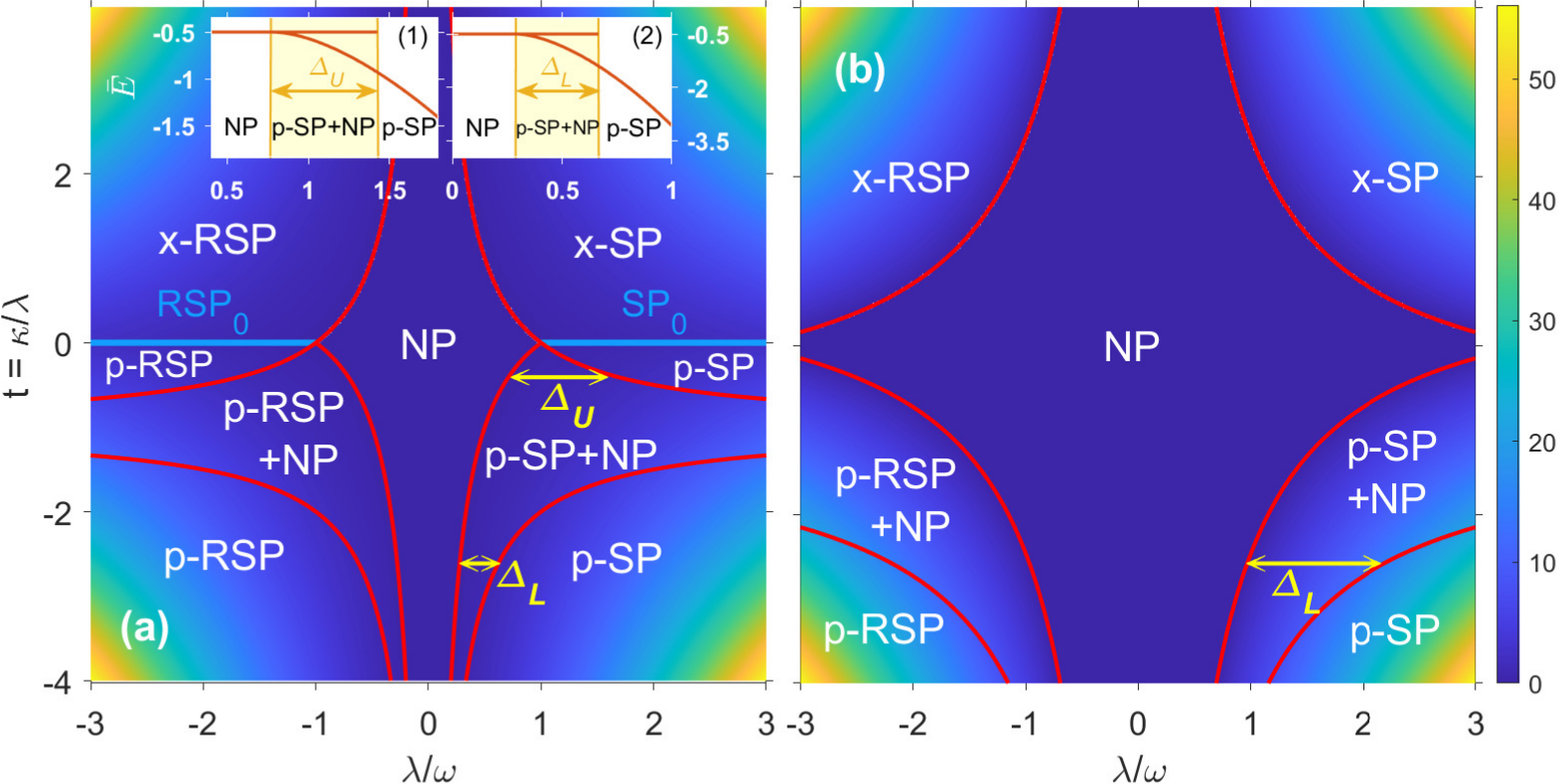}
  \caption{(Color online) The mean photon number $\langle a^\dagger a \rangle /N$ as a function of the collective atom-photon coupling strength $\lambda$ and the ratio $t=\kappa/\lambda$ at resonance (panel (a) $\Omega = \omega$) and off resonance (panel (b) $\Omega = 12\omega$), respectively. The blue solid lines represent the phases of the TC model, (R)SP$_0$. And they also indicate the critical boundaries of the different phases together with the red solid lines. $\Delta_U$ or $\Delta_L$ represents the width of $p$-(R)SP+NP under any negative fixed ratio $t$, where $t\in(-1,0)$ or $t<-1$, respectively. Insets (1) and (2) depict the scaled ground-state energy $\bar{E}$ as a function of $\lambda$ at resonance for $t=-0.3$ and $t=-2.5$, respectively.}
  \label{fig_PhaseDiagram}
\end{figure}

In Table \ref{table_phase2}, we present the analytical results of the mean photon number $\langle a^\dagger a \rangle /N $, the scaled atom population $\langle J_z \rangle /N$, and the scaled atom dipole moments $\langle J_x\rangle /N $ and $\langle J_y \rangle /N $, as well as the phase differences $\theta$ and $\eta$ in different phases of the system for nonzero $\kappa$ (i.e., $t\neq 0$). Although these results are derived under the situation of $t\neq 0$ and $\pm 1$, they are completely consistent with the existing results in the cases of $t=\pm 1$. To further understand the change of the ground-state behavior of the system versus $t$, it is convenient to plot the critical boundaries of the phases in the system. In Fig.~\ref{fig_PhaseDiagram}, we plot the mean photon number $\langle a^\dagger a \rangle /N$ against $\lambda$ and $t$ at resonance (panel (a) $\Omega=\omega$) and off resonance (panel (b) $\Omega=12\omega$), respectively. All critical boundaries, indicated by the red solid lines, are derived analytically through Eqs. (\ref{stable_NP_general_t}) and (\ref{stable_SP_general_t}) in \ref{App_A}. Remarkably, the critical boundaries of NP are symmetric with respect to $t=-1$, while those of SPs are symmetric about $t=0$. This result is consistent with Eqs. (\ref{stable_NP_general_t}) and (\ref{stable_SP_general_t}). Clearly, the coexisting phases depend on the sign of $t$ and can only occur when $t<0$. Additionally, the regions of NP and the two sign-dependent coexisting phases expand as $\Omega$ increases. Specifically, the width of the $p$-SP+NP or $p$-RSP+NP, given by $\Delta_L=2\sqrt{\omega\Omega}/(t^2-1)$ for a fixed $t<-1$ and $\Delta_U = 2t\sqrt{\omega\Omega}/(t^2-1)$ for a fixed $t\in (-1,0)$, gradually widens as $\Omega$ increases (see Fig. \ref{fig_PhaseDiagram}).

As mentioned above, the ground-state stability indicates that the sign-dependent coexisting phases occur naturally. To provide further insight into the $p$-SP/RSP+NP, in the insets (1) and (2) of Fig. \ref{fig_PhaseDiagram} we plot the scaled ground-state energy $\bar{E}$ as a function of $\lambda$ at resonance for different $t$. It is found that $\bar{E}$ of the SP is lower than that of the NP throughout the coexistence regions, confirming that the SP is the global ground state while the NP is metastable. Physically, although a gauge transformation can map the anti-Dicke model ($t=-1$) to the standard Dicke model ($t=1$) and preserve the energy spectrum, it does not preserve the stability of local minima in the free-energy landscape. The stability of these local minima, which governs the metastable dynamics, is sensitive to the sign of $t$. Consequently, for $t<0$, the free energy exhibits two local minima (NP and SP), giving rise to the sign-dependent coexisting phases; for $t>0$, no such metastable NP persists beyond criticality.

This sign dependence originates from the interplay between the co- and counterrotating coupling terms. When $t>0$, the two terms cooperate and jointly destabilize the NP at a single critical coupling, producing a continuous superradiant transition with no metastable NP. When $t<0$, the two terms have opposite signs and partially cancel each other's destabilizing effect on the NP, which weakens the effective drive that pushes the system out of the NP. As a result, the NP remains locally stable up to a significantly larger coupling $|\lambda_{c,\mathrm{NP}}|=\sqrt{\omega\Omega}/|1+t|$ (which diverges as $t\to -1$, where the NP remains locally stable at any finite $\lambda$). Meanwhile, the superradiant phase still emerges at $|\lambda_{c,p}|=\sqrt{\omega\Omega}/|1-t|$, which for $t<0$ is smaller than $|\lambda_{c,\mathrm{NP}}|$. The separation between these two critical boundaries creates the coexistence window, whose width grows as $t\to-1$. This competition mechanism is analogous to the interplay between imbalanced coupling channels that drives first-order transitions and metastability in cavity self-organization experiments~\cite{Leonard2017Mar,Li2021Mar}.

\begin{figure}[t]
  \centering\includegraphics[width = 1.0\linewidth]{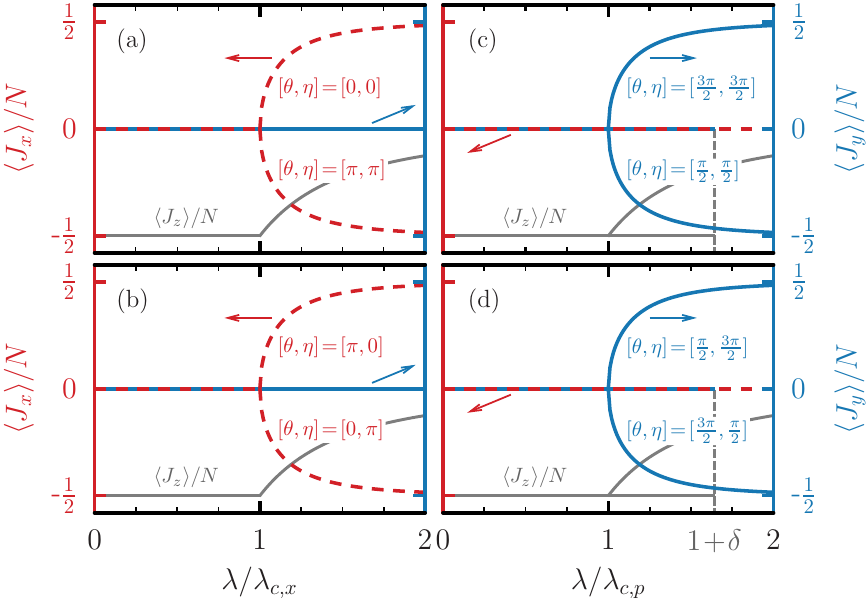}
  \caption{(Color online) The scaled atom dipole moments $\langle J_x \rangle /N$ (red dashed lines) and $\langle J_y \rangle /N$ (blue solid lines) versus $\lambda/\lambda_c$. When $\lambda/\lambda_c >1$, the system transitions into four different phases: (a) $x$-SP, (b) $x$-RSP, (c) $p$-SP+NP, and (d) $p$-RSP+NP, respectively. For comparison, the scaled atom population $\langle J_z \rangle /N$ (gray solid lines) is also given in each panel. The gray dashed lines at $\lambda/\lambda_{c,p} = 1 + \delta$ in panels (c) and (d) indicate the critical points of the sign-dependent coexisting phases vanishing. In each panel, the upper or lower branches of $\langle J_x \rangle /N$ or $\langle J_y \rangle /N$ correspond to different combinations of the phase differences $[\theta,\eta]$ of the corresponding phases.}
  \label{fig_JxJy}
\end{figure}

Figure \ref{fig_JxJy} shows the transitions of the scaled atom dipole moments $\langle J_x \rangle /N$ and $\langle J_y \rangle /N$ with respect to $\lambda/\lambda_c$. When $t>0$, the system undergoes a QPT from NP to (a) $x$-SP with $\lambda>0$ or (b) $x$-RSP with $\lambda<0$ at the critical point $\lambda = \lambda_{c,x}$. In the panel (a), the phase differences $[\theta,\eta]=[0,0]$ and $[\pi,\pi]$ correspond to the upper and lower branches of $\langle J_x \rangle /N$, while in the panel (b), $[\theta,\eta]=[\pi,0]$ and $[0,\pi]$ correspond to the upper and lower branches of $\langle J_x \rangle /N$ and the \textit{reversal} behavior of the phase differences occur. When $t<0$, the system first undergoes a QPT from NP to $p$-SP/RSP+NP at $\lambda=\lambda_{c,p}$, and then to $p$-SP/RSP at $\lambda=\lambda_{c,p}(1+\delta)$ with (c) $\lambda>0$ and (d) $\lambda<0$, respectively. Here the dimensionless parameter $\delta$ is determined by $\delta = \Delta_{L}/\lambda_{c,p} = -2/(1+t)$ when $t<-1$, or $\delta=\Delta_{U}/\lambda_{c,p} = -2t/(1+t)$ when $t\in(-1,0)$. And, in panel (c), the phase differences $[\theta,\eta] = [3\pi/2,3\pi/2]$ and $[\pi/2,\pi/2]$ correspond to the upper and lower branches of $\langle J_y \rangle/N$, respectively. In the same way, in panel (d) the phase differences $[\theta,\eta]$ are also reversed. 

\begin{figure}[b]
  \centering\includegraphics[width = 0.8\linewidth]{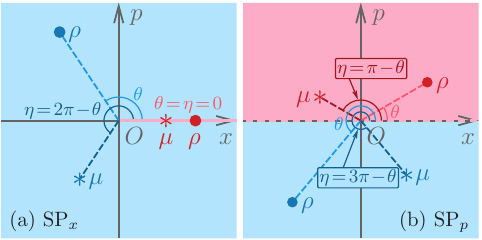}
  \caption{(Color online) Complex amplitudes in the abstract position-momentum representation of SP in the case of zero $\lambda$. (a) SP$_x$ ($\kappa>0$), in which the phase differences $\theta=\eta=0$ (the light red line), or $\theta+\eta=2\pi$ (the light blue region). (b) SP$_p$ ($\kappa<0$), in which $\theta + \eta =\pi$ (the light red region and the $x$ axis) or $3\pi$ (the light blue region). Here, the phase differences $\theta$ and $\eta$ are axially symmetric concerning the $x$- and $p$-axis, respectively. They behave not only differently from each other, but also very differently from those in SP$_0$/RSP$_0$ and $x$/$p$-SP/RSP.}
  \label{fig_zerolambda}
\end{figure}

Finally, for the exceptional case of $\lambda=0$, which has been addressed experimentally \cite{Zhiqiang2017Apr}, the Hamiltonian (\ref{H_0}) reduces to $H=\omega a^{\dagger}a+\Omega J_{z}+(\kappa /\sqrt{N})(a^{\dagger }J_{+}+aJ_{-})$. It is evident from Eqs. (\ref{MidEqsC}) and (\ref{MidEqsD}) that $\sin (\theta +\eta )=0$ for the SP. These two equations can therefore be fulfilled by taking $\theta =\eta =0$, or $\theta +\eta \in \{\pi, 2\pi, 3\pi \} $. And, according to Eq. (\ref{HM_m4}), the corresponding $\zeta_{+}$ would then be $|\kappa| $ for the stable SP since $\mathcal{M}^\prime = 2\rho\mu\mathcal{C} \geqslant 0$. Obviously, $\zeta _{+}$ arising from the different phase differences $\theta $ and $\eta $ will directly influence the ground-state properties of the system. Fig. \ref{fig_zerolambda} illustrated the complex amplitudes in the abstract position-momentum representation of SP in this case. When $\kappa>0$ (see Fig. \ref{fig_zerolambda} (a)), the phase differences $\theta=\eta=0$ (the light red line), or $\theta+\eta=2\pi$ (the light blue region). When $\kappa<0$ (see Fig. \ref{fig_zerolambda} (b)), $\theta + \eta =\pi$ (the light red region and the $x$ axis) or $3\pi$ (the light blue region). It is found that the phase differences $\theta$ and $\eta$ are axially symmetric about the $x$- or $p$-axis when $\kappa \gtrless 0$. Consequently, we name them SP$_x$ and SP$_p$, respectively. The ground-state properties in these two SPs behave very differently, and the symmetry breaking may occur in different ways as well.

Similar to the case of the TC model, it is easy to find $[\rho, \mu^2]=[0,0]$ (NP) or $[\frac{|\kappa|}{2\omega}\sqrt{1-\omega^2\Omega^2/\kappa ^{4}},$ $\frac{1}{2}(1-\omega\Omega/\kappa ^{2})]$ (SP). And it can be proved that the stability regions of different phases would also be determined in terms of the inequalities (\ref{stable_12_2}) and (\ref{stable_34_2}) in \ref{App_A}. In particular, we just need to substitute $\zeta_+ = |\kappa|$ into (\ref{stable_12_2}) and obtain
\begin{equation}
  \mu^2 \geqslant \frac{\kappa^2-\omega\Omega}{6\kappa^2}.
\end{equation}
Therefore, the stability region of NP is determined by $|\kappa| \leqslant \kappa_c$ and that of SP is $|\kappa| \geqslant \kappa_c$ with $\kappa_c = \sqrt{\omega\Omega}$ being the critical points.

In the fields of quantum control and quantum devices, rich critical behaviors are of great importance. For example, in quantum batteries, the maximum stored energy undergoes significant changes at the critical point of the QPT \cite{Dou2022Mar}. The novel spontaneous polarization phases and their superradiant phase transitions, which we have identified in the current imbalanced Dicke model, may play even more crucial roles in these areas \cite{Ferraro2018Mar, Wang2024Oct}.

\section{Symmetries of the system}
\label{Sec_Symmetry}

\begin{figure}[t]
  \centering\includegraphics[width = 1.0\linewidth]{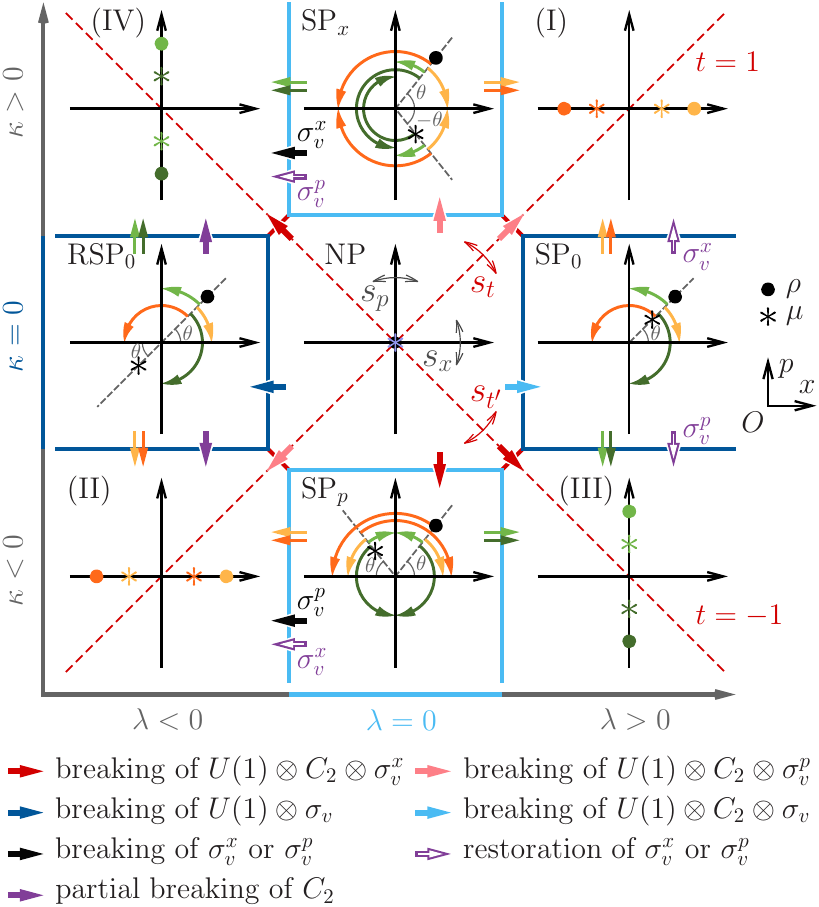}
  \caption{(Color online) Symmetry diagram of the system in the $\lambda$-$\kappa$ plane. The reflections $s_x$ and $s_p$ ($s_t$ and $s_{t^\prime}$), which fix the axes $x$ and $p$ (the lines $t=\pm 1$), generate the Coxeter group $W$ ($W^\prime$), respectively. Symmetries concerning $W$ and $W^\prime$ in different phases are detailed in Table \ref{table_symmetry}. The arrows in the legend represent the breaking, partial breaking, or restoration of the symmetries. Specifically, the light blue arrow (NP $\to$ SP$_0$) indicates complete breaking of all symmetries; the dark blue arrow (NP $\to$ RSP$_0$) indicates preservation of the $C_2$ symmetry; the three light red arrows (NP $\to$ SP$_x$/$x$-SP/$x$-RSP) and the three dark red arrows (NP $\to$ SP$_p$/$p$-SP/$p$-RSP) indicate preservation of $\sigma_v^x$ and $\sigma_v^p$, respectively. The purple hollow arrows (SP$_0$ $\to$ $x$-SP/$p$-SP) indicate restoration of $\sigma_v^x/\sigma_v^p$; the purple solid arrows (RSP$_0$ $\to$ $x$-RSP/$p$-RSP) indicate partial breaking of $C_2$ into $\sigma_v^x/\sigma_v^p$; the black arrows together with the horizontal purple hollow arrows (SP$_x$/SP$_p$ $\to$ $p$-RSP/$x$-RSP) indicate the exchange between $\sigma_v^x$ and $\sigma_v^p$: the black arrows break one symmetry while the purple hollow arrows restore the other. The yellow arc arrows and green arc arrows illustrate the behaviors of $\theta$ and $\eta(\theta)$ when the system transforms from SP$_0$/RSP$_0$ and SP$_x$/SP$_p$ into $x$/$p$-SP/RSP. Note that to demonstrate the behaviors of $\theta$ and $\eta(\theta)$ in SP$_0$/RSP$_0$ (SP$_x$/SP$_p$), we enlarged the region where $\kappa=0$ ($\lambda =0$), indicated by the dark (light) blue line on the $\kappa$-axis ($\lambda$-axis), respectively.}
  \label{fig_symmetry}
\end{figure}

To find the symmetries of the Hamiltonian (\ref{H_system}), we first consider the unitary transformation $\mathcal{U}=\mathcal{U}_\theta \circ \mathcal{U}_\eta^\prime$ with $\mathcal{U}_\theta = \exp{(\mathrm{i}\theta a^\dagger a)}$ and $\mathcal{U}_\eta^\prime = \exp{[\mathrm{i}\eta (J_z + N/2)]}$ acting on the field and matter sectors respectively. Remarkably,
\begin{equation}
  \mathcal{U} (a, a^\dagger, J_+, J_-) \mathcal{U}^\dagger = (a \mathrm{e}^{-\mathrm{i}\theta}, a^\dagger \mathrm{e}^{\mathrm{i}\theta}, J_+ \mathrm{e}^{\mathrm{i}\eta}, J_- \mathrm{e}^{-\mathrm{i}\eta }). \label{UnitaryTrans}
\end{equation}
Noticing that the nonlinear atom-photon interaction $U$ does not alter the symmetry of the system, we can directly discuss the symmetry based on the imbalanced Dicke Hamiltonian (\ref{H_0}). By substituting Eq. (\ref{UnitaryTrans}) into Hamiltonian (\ref{H_0}),
\begin{eqnarray}
  \mathcal{U}H\mathcal{U}^\dagger \!&=&\! \omega a^\dagger a \!+\! \Omega J_z \!+\! \frac{\lambda}{\sqrt{N}}[\mathrm{e}^{\mathrm{i}(\theta-\eta)}a^\dagger J_- \!+\! \mathrm{e}^{-\mathrm{i}(\theta-\eta)} a J_+] \notag \\
              && + \frac{\kappa}{\sqrt{N}}[\mathrm{e}^{\mathrm{i}(\theta+\eta)} a^\dagger J_+ \!+\! \mathrm{e}^{-\mathrm{i}(\theta+\eta)} a J_-].
  \label{H_U}
\end{eqnarray}
Considering the physical nature of the phase differences, for simplicity, we henceforth redefine $\theta, \eta \in [-\pi, \pi]$ in SP$_0$ and SP$_x$; whereas $\theta \in [-\pi, \pi]$ and $\eta \in [0, 2\pi]$ in RSP$_0$ and SP$_p$, respectively. In the TC model ($\kappa=0$), the counter-rotating terms $a^\dagger J_+$ and $aJ_-$ vanish. Therefore, the system is invariant under the transformation $\mathcal{U}_\theta \circ \mathcal{U}^\prime_{\theta}$ ($\lambda>0$) or $\mathcal{U}_\theta \circ \mathcal{U}^\prime_{\theta+\pi}$ ($\lambda<0$), and possesses a continuous $U(1)$ symmetry \cite{Baksic2014}. Similarly, for the case of $\lambda=0$, the rotating terms $a J_+$ and $a^\dagger J_-$ are neglected \cite{Zhiqiang2017Apr}. The field and matter sectors are invariant under the transformation $\mathcal{U}_\theta \circ \mathcal{U}^\prime_{-\theta}$ ($\kappa>0$) or $\mathcal{U}_\theta \circ \mathcal{U}^\prime_{-\theta+\pi}$ ($\kappa<0$), and also features a continuous $U(1)$ symmetry. As Fig. \ref{fig_symmetry} shows, when $\lambda=0$ ($\kappa=0$), the spontaneous breaking of $U(1)$ leads to a QPT from NP to SP$_0$ (SP$_x$) or RSP$_0$ (SP$_p$), respectively.

It is noteworthy that we identified an additional symmetry concerning the phase differences, $\theta$ and $\eta$, which can be characterized by a discrete Coxeter group, $W = A_1\oplus A_1$ \cite{GroveCoxeter}. The group $W$ is generated by $\{s_x,s_p\}$ and determined by the relations
\begin{equation}
  (s_x s_x)^1 = (s_p s_p)^1 = (s_x s_p)^2 = (s_p s_x)^2 = e.
  \label{W_generator}
\end{equation}
Here $s_x$ and $s_p$ are two reflections in $\mathbb{R}^2$, fixing the axes $x$ and $p$ that make an angle $\pi/2$ as shown in Fig. \ref{fig_symmetry}. The Coxeter group $W=\{e, s_x, s_p, s_x s_p\}$ contains an identity symmetry $C_1$, two reflection symmetries $\sigma_v$s and a central symmetry $C_2$, which are independently established by four subgroups $\mathcal{I} = \{e\}$, $W_x=\{e, s_x\}$, $W_p = \{e, s_p\}$, and $\mathcal{C}_2 = \{e, s_x s_p\}$, respectively. Crucially, the symmetries determined by $W$ and its subgroups can be broken either simultaneously or independently. Notice that $C_1$ is an identity operator, providing a universal yet trivial transformation, and thus can be eliminated henceforth. The reflection symmetries $\sigma_v^{x}$ and $\sigma_v^p$ yield equivalent transformations:
\begin{subequations}
  \label{Sym_Tx_Tp}
  \begin{align}
  \mathcal{T}_x &:(x_a, p_a, J_x, J_y)  \rightarrow   (x_a, -p_a, J_x, -J_y), \\
  \mathcal{T}_p &:(x_a, p_a, J_x, J_y)  \rightarrow   (-x_a, p_a, -J_x, J_y).
  \end{align} 
\end{subequations}
The cyclic group $\mathcal{C}_2$, by acting with the projectors of the symmetric and antisymmetric representations, restores symmetry in a manner analogous to a discrete $\mathbb{Z}_2$ symmetry \cite{Castanos2010, Castanos2011May, Castanos2011} associated with the transformation $\mathcal{T}_x \circ \mathcal{T}_p$ \cite{Baksic2014}. Specifically, $p_a \rightarrow -p_a$ and $x_a \rightarrow -x_a$ represent spatial reflections $\mathcal{P}$ about the $x$-axis and $p$-axis, respectively. And $J_y \rightarrow -J_y$ implies a spatial reflection $\mathcal{P}$ through the $J_x$-$J_z$ plane; similarly, $J_x \rightarrow -J_x$ stands for a spatial reflection $\mathcal{P}$ through the $J_y$-$J_z$ plane. The spatial reflection operator $\mathcal{P}$ behaves exactly like the parity operator in our scenario \cite{Jin2016Feb}.

\begin{table}[t]
  \caption{Symmetries in different phases. The symbols $\checkmark$ and $\times$ indicate the presence and absence of the corresponding symmetries, respectively.}
  \label{table_symmetry}
  \tabcolsep=0.03\linewidth
  \begin{tabular*}{\linewidth}{l c c c c c c c c}
  \specialrule{0.08em}{3pt}{5pt}
    \multirow{2}{*}{Phase} & \multirow{2}{*}{$U(1)$} & \multicolumn{3}{c}{$W$} && \multicolumn{3}{c}{$W^\prime$}\\ \cline{3-5} \cline{7-9} \specialrule{0em}{1pt}{1pt}
            & ~ & $\sigma_v^x$ & $\sigma_v^p$ & $C_2$ && $\sigma_v^t$ & $\sigma_v^{t^\prime}$ & $C_2^\prime$ \\ \specialrule{0.05em}{1pt}{5pt}
    NP      & $\checkmark$ & $\checkmark$ & $\checkmark$ & $\checkmark$ && $\checkmark$ & $\checkmark$ & $\checkmark$ \\ \specialrule{0em}{3pt}{3pt}
    SP$_0$  & $\times$     & $\times$     & $\times$     & $\times$     && $\times$     & $\times$     & $\times$     \\ \specialrule{0em}{1pt}{1pt}
    RSP$_0$ & $\times$     & $\times$     & $\times$     & $\checkmark$ && $\times$     & $\times$     & $\times$     \\ \specialrule{0em}{1pt}{1pt}
    SP$_x$  & $\times$     & $\checkmark$ & $\times$     & $\times$     && $\times$     & $\times$     & $\times$     \\ \specialrule{0em}{1pt}{1pt}
    SP$_p$  & $\times$     & $\times$     & $\checkmark$ & $\times$     && $\times$     & $\times$     & $\times$     \\ \specialrule{0em}{3pt}{3pt}
    $x$-SP  & $\times$     & $\checkmark$ & $\times$     & $\times$     && $\checkmark$ & $\times$     & $\times$     \\ \specialrule{0em}{1pt}{1pt}
    $x$-RSP & $\times$     & $\checkmark$ & $\times$     & $\times$     && $\checkmark$ & $\times$     & $\times$     \\ \specialrule{0em}{1pt}{1pt}
    $p$-SP  & $\times$     & $\times$     & $\checkmark$ & $\times$     && $\times$     & $\checkmark$ & $\times$     \\ \specialrule{0em}{1pt}{2pt}
    $p$-RSP & $\times$     & $\times$     & $\checkmark$ & $\times$     && $\times$     & $\checkmark$ & $\times$     \\ \specialrule{0.08em}{3pt}{0pt}
  \end{tabular*}
\end{table}

Configurations of the phase differences $[\theta,\eta{(\theta)}]$ and the symmetries concerning $W$ in different phases are illustrated in Fig. \ref{fig_symmetry} and Table \ref{table_symmetry}. All symmetries are preserved in the NP but broken in the SP$_0$, as indicated by the light blue arrow in Fig. \ref{fig_symmetry}. In RSP$_0$, SP$_x$, and SP$_p$, the $C_2$, $\sigma_v^x$, and $\sigma_v^p$ symmetries are respectively preserved, as indicated by the dark blue arrow and the two longitudinal red arrows in Fig. \ref{fig_symmetry}. These symmetry breakings are accompanied by QPTs from NP to SP$_0$/RSP$_0$ or SP$_{x}$/SP$_{p}$, respectively.

Figure \ref{fig_symmetry} also illustrates the changes of the phase differences $\theta$ and $\eta$ when the system transforms from SP$_0$/RSP$_0$ or SP$_{x}$/SP$_{p}$ to $x$/$p$-SP/RSP. Specifically, when $\theta$ and $\eta(\theta)$ vary along the (dark) yellow arc arrows or the (dark) green arc arrows, they will be locked into $\{0, \pi\}$ or $\{\pi/2,3\pi/2\}$ independently. According to Table \ref{table_symmetry}, $x$-SP/RSP and $p$-SP/RSP possess the $\sigma_v^x$ and $\sigma_v^p$ symmetries, respectively, and thus remain invariant under the transformations $\mathcal{T}_x$/$\mathcal{T}_p$, respectively. Consequently, the QPT from SP$_0$ to $x$-SP/$p$-SP is accompanied by the restoration of the $\sigma_v^x$/$\sigma_v^p$ symmetry, as well as the corresponding transformation $\mathcal{T}_x$/$\mathcal{T}_p$, respectively (see the two longitudinal purple hollow arrows in Fig. \ref{fig_symmetry}). Similarly, the QPT from RSP$_0$ to $x$-RSP/$p$-RSP is accompanied by a partial breaking of $C_2$ symmetry, specifically into $\sigma_v^x$/$\sigma_v^p$, respectively. Thus the invariance of $\mathcal{T}_x$/$\mathcal{T}_p$ is also broken. (see the purple arrows in Fig. \ref{fig_symmetry}). The QPT from SP$_x$/SP$_p$ to $p$-RSP/$x$-RSP involves the breaking of $\sigma_v^x$/$\sigma_v^p$ and the simultaneous restoration of $\sigma_v^p$/$\sigma_v^x$. This transformation also applies to $\mathcal{T}_x$/$\mathcal{T}_p$ (see the two black arrows and the two horizontal purple hollow arrows in Fig. \ref{fig_symmetry}). However, the QPT from SP$_x$/SP$_p$ to $x$-SP/$p$-SP does not break or restore any symmetry concerning the Coxeter group $W$. 

On the other hand, it is well established that the standard Dicke model ($t=1$) breaks a discrete $\mathbb{Z}_2$ symmetry and undergoes a second-order phase transition in the thermodynamic limit \cite{Emary2003a, Emary2003}. When $\lambda>0$, since the phase differences of the cavity mode and the atomic spin excitation are locked at 0 or $\pi$ simultaneously \cite{Baumann2011}, the parity of the excitation number operator $\Pi_s=\mathcal{U}_{\pi}\circ\mathcal{U}^\prime_{\pi}=\exp[\mathrm{i}\pi(a^\dagger a + J_z + N/2)]$ is conserved and transforms the operators by $\Pi_s:(a, a^\dagger, J_{\pm})\rightarrow -(a, a^\dagger, J_{\pm})$. However, when $\lambda<0$, the field and matter sectors are invariant under the transformation $\mathcal{T}_\lambda \circ \mathcal{T}_\kappa \circ \Pi_{s}^{R}$ with $\Pi_{s}^{R} = \mathcal{U}_0 \circ \mathcal{U}^\prime_\pi$ or $\mathcal{U}_\pi \circ \mathcal{U}^\prime_0$ and another discrete $\mathbb{Z}_2$ symmetric mapping $\mathcal{T}_{\lambda(\kappa)}:\lambda(\kappa) \rightarrow -\lambda(-\kappa)$ \cite{Liu2017}. For the anti-Dicke model ($t=-1$), the phase differences are locked to either $\{\pi/2, 3\pi/2\}$. It is noted that when $\lambda>0$, the model is invariant under the transformation $\mathcal{T}_\kappa \circ \mathcal{U}_{\pi/2} \circ \mathcal{U}^\prime_{\pi/2}$ or $\mathcal{T}_\kappa \circ \mathcal{U}_{3\pi/2} \circ \mathcal{U}^\prime_{3\pi/2}$; when $\lambda<0$, it is invariant under $\mathcal{T}_\lambda \circ \mathcal{U}_{\pi/2} \circ \mathcal{U}^\prime_{3\pi/2}$ or $\mathcal{T}_\lambda \circ \mathcal{U}_{3\pi/2} \circ \mathcal{U}^\prime_{\pi/2}$. Therefore, as indicated by the four diagonal red arrows in Fig. \ref{fig_symmetry}, the QPTs from NP to SPs are always accompanied by the partial breaking of $\mathbb{Z}_2$ ($\sigma_v^x$ or $\sigma_v^p$). The above results are also applicable to the general imbalanced scenario $t\neq \pm1$. 

In addition to the symmetries related to the Coxeter group $W$, the imbalanced model also exhibits further discrete symmetries associated with two other transformations:
\begin{subequations}
  \label{Sym_V}
  \begin{align}
  \mathcal{V} &:(\eta,\lambda,\kappa) \rightarrow  (-\eta, \kappa, \lambda), \\
  \mathcal{V}^\prime &:(\eta,\lambda,\kappa) \rightarrow  (-\eta+\pi, -\kappa, -\lambda).
  \end{align}
\end{subequations}
Both $\mathcal{V}$ and $\mathcal{V}^\prime$ combine a time-reversal symmetry $\mathcal{T}$ \cite{Jin2016Feb} and a parameter exchange symmetry $\mathcal{T}_\mathrm{ex}$ \cite{Stitely2020Jul}, leaving the Hamiltonian (\ref{H_0}) invariant. 

Specifically, under the transformation $\mathcal{V}$, the change in sign of $\eta$ suggests a time-reversal symmetry $\mathcal{T}$ \cite{Jin2016Feb} since it behaves as a synthetic magnetic flux of the matter field \cite{Wilczek1982Apr, Jin2017Jan, Barbiero2019Oct, Chiacchio2023Sep}; and the exchange of $\lambda$ and $\kappa$ implies a parameter reflection $\mathcal{T}_\mathrm{ex}$ corresponding to the diagonal line $t=1$ in the $\lambda$-$\kappa$ plane. Physically, the parameter exchange symmetry is associated with the invariance under the exchange of the rotating frame about which the co- and counterrotating terms oscillate \cite{Stitely2020Jul}. The transformation $\mathcal{V}$ seemingly leads to the exchanges of $p$-SP $\rightleftharpoons$ $p$-RSP, SP$_0$ $\rightleftharpoons$ SP$_x$, and SP$_p$ $\rightleftharpoons$ RSP$_0$ in the symmetry diagram, but leaves them unchanged. This conclusion can be addressed directly by applying $\mathcal{V}$ to Table \ref{table_phase2} and the symmetry diagram in Fig. \ref{fig_symmetry}.  

Similarly, under $\mathcal{V}^\prime$, the transformation of $(\lambda,\kappa) \rightarrow (-\kappa,-\lambda)$ implies a parameter reflection corresponding to the antidiagonal line $t=-1$ in the $\lambda$-$\kappa$ plane. This reflects an invariance under the exchange and reversal of the rotating frame physically. Moreover, the additional $\pi$ in $-\eta$ is attributed to the $\mathbb{Z}_2$ symmetry, ensuring that the system remains invariant under the transformation. Notably, $\mathcal{V}^\prime$ leads to the exchanges of $x$-SP $\rightleftharpoons$ $x$-RSP, SP$_0$ $\rightleftharpoons$ SP$_p$, and SP$_x$ $\rightleftharpoons$ RSP$_0$ in the symmetry diagram shown in Fig. \ref{fig_symmetry}. It is also consistent with the results presented in Table \ref{table_phase2}.

In fact, the transformations $\mathcal{V}$ and $\mathcal{V}^\prime$ are equivalent to another Coxeter group $W^\prime = A_1\oplus A_1$ in the $\lambda$-$\kappa$ plane. The group $W^\prime$ is generated by $\{s_t,s_{t^\prime}\}$, where $s_t$ and $s_{t^\prime}$ are two reflections in $\mathbb{R}^2$ associated with the lines $t=\pm 1$, as shown in Fig. \ref{fig_symmetry}. And the relations similar to Eq. (\ref{W_generator}) are also satisfied:
\begin{equation}
  (s_t s_t)^1 = (s_{t^\prime} s_{t^\prime})^1 = (s_t s_{t^\prime})^2 = (s_{t^\prime} s_t)^2 = e.
  \label{Wprime_generator}
\end{equation}
Different from the Coxeter group $W$, which primarily describes the reflections of the phase differences $\theta$ and $\eta$ in the abstract position-momentum representation, $W^\prime$ generally concerns the changes of the function $\zeta_+$ and the phase difference $\eta$ of the SPs in the $\lambda$-$\kappa$ plane. The corresponding symmetries regarding $W^\prime$ in different phases are listed in Table \ref{table_symmetry}. According to $W^\prime$, we can derive the invariance in detail, such as
(i) $p$-SP $\stackrel{s_t}{\rightleftharpoons}$ $p$-RSP with $\zeta_+=\lambda-\kappa \stackrel{s_t}{\rightleftharpoons} \kappa-\lambda$ and $[\theta,\eta]=[\pi/2,\pi/2] \stackrel{s_t}{\rightleftharpoons} [\pi/2,3\pi/2]$ or $[3\pi/2,3\pi/2] \stackrel{s_t}{\rightleftharpoons} [3\pi/2,\pi/2]$;
(ii) $x$-SP $\stackrel{s_{t^\prime}}{\rightleftharpoons}$ $x$-RSP with $\zeta_+=\lambda+\kappa \stackrel{s_{t^\prime}}{\rightleftharpoons} -\kappa-\lambda$ and $[\theta,\eta]=[0,0] \stackrel{s_{t^\prime}}{\rightleftharpoons} [0,\pi]$ or $[\pi,\pi] \stackrel{s_{t^\prime}}{\rightleftharpoons} [\pi,0]$;
(iii) SP$_0$ $\stackrel{s_t}{\rightleftharpoons}$ SP$_x$ $\stackrel{s_{t^\prime}}{\rightleftharpoons}$ RSP$_0$ $\stackrel{s_t}{\rightleftharpoons}$ SP$_p$ $\stackrel{s_{t^\prime}}{\rightleftharpoons}$ SP$_0$ with $\zeta_+ = \lambda \stackrel{s_t}{\rightleftharpoons} \kappa \stackrel{s_{t^\prime}}{\rightleftharpoons} -\lambda \stackrel{s_t}{\rightleftharpoons} -\kappa \stackrel{s_{t^\prime}}{\rightleftharpoons} \lambda$ and $[\theta,\eta] = [\theta,\theta] \stackrel{s_t}{\rightleftharpoons} [\theta,-\theta] \stackrel{s_{t^\prime}}{\rightleftharpoons} [\theta,\theta+\pi] \stackrel{s_t}{\rightleftharpoons} [\theta,-\theta+\pi] \stackrel{s_{t^\prime}}{\rightleftharpoons} [\theta,\theta]$;
and the quadratic reflections SP$_0$ $\stackrel{s_t s_{t^\prime}}{\rightleftharpoons}$ RSP$_0$, SP$_x$ $\stackrel{s_t s_{t^\prime}}{\rightleftharpoons}$ SP$_p$, and so on.

Furthermore, by considering the transformations (\ref{Sym_Tx_Tp}) and (\ref{Sym_V}) together, the combined transformation results in a discrete parity-time ($\mathcal{PT}$) symmetry. It is found that, in the imbalanced Dicke model, the SPs are $\mathcal{PT}$ invariant \cite{Chiacchio2023Sep}.

\section{Conclusion and outlook}
\label{Sec_Conclusion}
In summary, we have investigated the ground-state properties and QPTs of the ultracold atomic ensemble in a Raman-assisted cavity. The system is governed by an imbalanced Dicke Hamiltonian. We have revealed several spontaneous polarized phases with distinct behaviors of phase differences $\theta$ and $\eta$ , such as $x$/$p$-SP/RSP, SP$_0$/RSP$_0$, and SP$_{x}$/SP$_{p}$, depending on the function $\zeta_+(\theta,\eta)$.
Crucially, we identify two sign-dependent coexisting phases ($p$-SP/RSP+NP), which emerge exclusively when the coupling ratio $t = \kappa/\lambda < 0$. In this coexistence region, the $p$-SP/RSP is the global ground state, while the NP remains a persistent metastable state. This phenomenon is absent in the standard Dicke model, where the NP is unstable beyond the critical point.
The rich phase diagrams, characterized by the order parameters, including the mean photon number $\langle a^{\dagger}a\rangle / N$, the scaled atom population $\langle J_z \rangle /N$, and the scaled atom dipole moments $\langle J_x \rangle /N$ and $\langle J_y \rangle /N$, provided a wealth of QPTs and indicated rich symmetries in the system. Moreover, we have demonstrated the breaking, partial breaking, and restoration of the symmetry of the system. It was discovered that, apart from the well-known $U(1)$ and $\mathbb{Z}_2$ symmetries, the system also featured several additional symmetries in detail, such as the reflection symmetries $\sigma_v$s, the central symmetry $C_2$ in the abstract position-momentum representation, the $\mathcal{PT}$ symmetry and the parameter exchange symmetry $\mathcal{T}_\mathrm{ex}$ in the parameter space. These additional symmetries are equivalent to two Coxeter groups $W$ and $W^\prime$. The invariance of the system can be pursued in detail relying on the subgroups of $W$ and $W^\prime$. These predictions are within experimental reach: the proposed model has been realized with $^{87}$Rb atoms in an optical cavity~\cite{Baden2014, Zhiqiang2017Apr}, and the order parameters and phase-difference configurations are accessible via heterodyne measurement of the intracavity field~\cite{Baumann2011, Li2021Mar, Esslinger2022Aug, Guo2019MayPRL}.

It is important to note that our analysis is based on the closed-system Hamiltonian. In open systems, cavity dissipation can alter the critical behavior, particularly suppressing the continuous $U(1)$ symmetry breaking along the TC line~\cite{Soriente2018May}. However, the $\mathbb{Z}_2$ symmetry-breaking transitions characteristic of the Dicke-like regimes (which dominate our imbalanced parameter space) are more robust against dissipation. Furthermore, the sign-dependent coexistence phases we identified correspond to metastable minima in the free-energy landscape, which may manifest as hysteresis loops in driven-dissipative settings, retaining their physical significance. Moreover, the symmetry classification governed by Coxeter groups $W$ and $W^\prime$ can also serve as a foundational framework for exploring symmetry-protected phenomena in driven-dissipative regimes. Their subgroups may remain invariant under certain dissipation channels, potentially leading to robust symmetry-protected phases.

\section*{Acknowledgments}
  This work is supported by the National Natural Science Foundation of China (NSFC) under Grant Nos. 11404216, 12074342, 12174461, 12234012, 12334012, 52327808, the Zhejiang Provincial Natural Science Foundation (ZJNSF) under Grant Nos. LY18A040004, LR22A040001, the National Key R\&D Program of China under grants Nos. 2021YFA1400900, 2021YFA0718300, 2021YFA1402100, and the Space Application System of China Manned Space Program.

\appendix

\section{Phase boundaries of the system}
\label{App_A}

Critical boundaries of the phases in the system are mainly determined by the Hessian matrix $\mathcal{M}$ (\ref{Hessian}) and its eigenvalues (\ref{HM_m}). Remarkably, since $\left( \mathcal{M}_{11}-\mathcal{M}_{22}\right) ^{2}+4\mathcal{M}_{12}^{2} \geqslant 0$, we have $m_2 \geqslant m_1 > 0$ in the stable phases, and thus $\min {\{m_1, m_3, m_4\}} = 0$ on the critical lines. As a result, the condition of $\{m_1, m_3, m_4\} \geqslant 0$ lead to a set of inequalities
\begin{eqnarray}
  \omega \Omega \left( 1-\mu ^{2}\right) ^{\frac{3}{2}}+\omega \rho \mu \zeta _{+}\left( 3-2\mu ^{2}\right) \geqslant \notag \\
  \zeta_{+}^{2}\left( 1-\mu ^{2}\right) ^{\frac{1}{2}}\left( 1-2\mu ^{2}\right) ^{2},   \label{stable_12}
\end{eqnarray}
and
\begin{subequations}
  \label{stable_34}
  \begin{align}
    m_{3} = 4\lambda \rho \mu \left( 1-\mu ^{2}\right) ^{\frac{1}{2}}\cos\left( \theta -\eta \right) \geqslant 0, \\
    m_{4} = 4\kappa \rho \mu \left( 1-\mu ^{2}\right) ^{\frac{1}{2}}\cos \left(  \theta +\eta \right) \geqslant 0.
  \end{align}
\end{subequations}

As mentioned in Sec. 4, Eqs.~(\ref{MidEqsA}) and~(\ref{MidEqsB}) result in
\begin{equation}
  \rho\mu = \frac{\zeta_{+}}{\omega}\mu^2 \sqrt{1-\mu ^{2}}.
  \label{rhomu}
\end{equation}
Plugging Eq. (\ref{rhomu}) into (\ref{stable_12}) then gives $(1-\mu^2)(6\zeta_+^2 \mu^2 - \zeta_+^2 +\omega\Omega) \geqslant 0$. Since $1-\mu^2>0$, it is found that
\begin{equation}
  \mu^2 \geqslant \frac{\zeta_+^2-\omega\Omega}{6\zeta_+^2}. \label{stable_12_2}
\end{equation}
Similarly, by plugging Eq. (\ref{rhomu}) into (\ref{stable_34}), we have
\begin{subequations}
  \label{stable_34_2}
  \begin{align}
    m_{3} = \frac{4\lambda \zeta_{+}}{\omega}\mu^2 \left( 1-\mu ^{2}\right) \cos\left( \theta -\eta \right) \geqslant 0, \\
    m_{4} = \frac{4\kappa \zeta_{+}}{\omega}\mu^2  \left( 1-\mu ^{2}\right) \cos \left(  \theta +\eta \right) \geqslant 0.
  \end{align}
\end{subequations}
Inequalities (\ref{stable_12_2}) and (\ref{stable_34_2}) would determine the stability regions of different phases of the system together. And the further specific results in different phases apparently depend on the values of $\mu^2$ as well as the phase differences $\theta$ and $\eta$. 

\subsection{Critical boundaries of NP}

When the system is in the NP, one obtains a trivial solution of $\rho =\mu =0$. Meanwhile, the function $\zeta _{+}=\lambda + \kappa $. Since the stability of the system in NP totally depends on $m_{1}$, by plugging these results into (\ref{stable_12_2}), we have
\begin{equation}
  |\lambda+\kappa| \leqslant \sqrt{\omega \Omega} \text{ or } |1+t| \leqslant \sqrt{\omega\Omega}/|\lambda|.
  \label{stable_NP}
\end{equation}

As a direct consequence of the inequality (\ref{stable_NP}), the stability region of NP for the TC model ($t=0$) is given by $|\lambda| \leqslant \sqrt{\omega\Omega}$; while for the standard Dicke model ($t=1$), it is subject to $|\lambda|\leqslant \sqrt{\omega\Omega}/2$; however, for the anti-Dicke model ($t=-1$), it becomes $0\leqslant \sqrt{\omega\Omega}$, which is always true in our circumstance. That is to say, the NP of the anti-Dicke model remains stable throughout. Generally, it is obtained 
\begin{subequations}
  \label{stable_NP_general}
  \begin{eqnarray}
    \kappa \leqslant & -\lambda +\sqrt{\omega\Omega}, &\text{ if }\lambda + \kappa \geqslant 0, \\
    \kappa \geqslant & -\lambda -\sqrt{\omega\Omega}, &\text{ if }\lambda + \kappa < 0,
  \end{eqnarray}
\end{subequations}
or
\begin{subequations}
  \label{stable_NP_general_t}
  \begin{eqnarray}
    t \leqslant &  \frac{\sqrt{\omega \Omega}}{\left\vert \lambda \right\vert }-1, &\text{ if }t\geqslant -1, \\
    t \geqslant &  -\frac{\sqrt{\omega \Omega}}{\left\vert \lambda \right\vert }-1, &\text{ if }t<-1.
  \end{eqnarray}
\end{subequations}

\subsection{Critical boundaries of SP}

When the system is in the SP, there exist nonzero coherences of the bosonic field and the spontaneous polarization of the collective spin. Thus we have $\mu ^{2}= \frac{1}{2} ( 1- {\omega \Omega }/{\zeta_{+}^{2}})$ and $\zeta_+ = \lambda \cos ( \theta -\eta ) + \kappa \cos( \theta +\eta )$ with $\lambda \cos ( \theta -\eta ) = |\lambda|$ and $\kappa \cos( \theta +\eta ) = |\kappa|$. The solution should satisfy the inequalities (\ref{stable_12_2}) and (\ref{stable_34_2}), then it is obtained that
\begin{equation}
  \zeta_+ \geqslant \sqrt{\omega\Omega}, \label{stable_SP_12}
\end{equation}
and
\begin{subequations}
  \label{stable_SP_34}
  \begin{eqnarray}
    m_{3} &=& \frac{\left\vert \lambda \right\vert \zeta _{+}}{\omega }\left( 1-\frac{\omega ^{2}\Omega ^{2}}{\zeta _{+}^{4}}\right) \geqslant 0, \\
    m_{4} &=& \frac{\left\vert \kappa \right\vert \zeta _{+}}{\omega }\left( 1-\frac{\omega ^{2}\Omega ^{2}}{\zeta _{+}^{4}}\right) \geqslant 0.
    \end{eqnarray}
\end{subequations}
Inequalities (\ref{stable_SP_34}) imply $\zeta_{+}\geqslant \sqrt{\omega \Omega }$ as well, which is fully in accord with (\ref{stable_SP_12}). This means that the phase boundaries of SP determined by (\ref{stable_12_2}) and (\ref{stable_34_2}) are completely identical.

Specifically, for the TC model ($t=0$), in which $\theta =\eta$ ($\lambda >0$) or $\eta \pm \pi $ ($\lambda <0$), we have $\zeta _{+}=\lambda \cos (\theta -\eta )= \vert \lambda \vert$, and the stability region of SP is given by $|\lambda| \geqslant \sqrt{\omega\Omega}$; while for the standard Dicke and anti-Dicke model ($t=\pm 1$), we have $\zeta_+ = |2\lambda|$, and the stability region of SP is given by $|\lambda|\geqslant \sqrt{\omega\Omega}/2$. However, for the other case of $t \neq 0,\pm 1$, since $\zeta _{+}=\vert \lambda +\kappa \vert$ ($x$-SP/RSP) and $\zeta _{+}=\vert \lambda -\kappa \vert $ ($p$-SP/RSP). According to (\ref{stable_SP_12}), it is obtained directly that $\left\vert \lambda +\kappa \right\vert \geqslant \sqrt{\omega \Omega }$ in $x$-SP/RSP; whereas $\left\vert \lambda -\kappa \right\vert \geqslant \sqrt{\omega \Omega }$ in $p$-SP/RSP.

It is generally found that
\begin{subequations}
  \label{stable_SP_general}
  \begin{eqnarray}
    \kappa \geqslant & -\lambda +\sqrt{\omega\Omega}, &\text{ if }\lambda>0 \text{ and } \kappa \geqslant 0, \\ 
    \kappa \leqslant & -\lambda -\sqrt{\omega\Omega}, &\text{ if }\lambda<0 \text{ and } \kappa \leqslant 0, \\ 
    \kappa \leqslant &  \lambda -\sqrt{\omega\Omega}, &\text{ if }\lambda>0 \text{ and } \kappa \leqslant 0, \\ 
    \kappa \geqslant &  \lambda +\sqrt{\omega\Omega}, &\text{ if }\lambda<0 \text{ and } \kappa \geqslant 0.    
  \end{eqnarray}
\end{subequations}
or
\begin{subequations}
  \label{stable_SP_general_t}
  \begin{eqnarray}
    t \geqslant &  \frac{\sqrt{\omega \Omega}}{\left\vert \lambda \right\vert }-1, &\text{ if }t\geqslant 0, \\
    t \leqslant & -\frac{\sqrt{\omega \Omega}}{\left\vert \lambda \right\vert }+1, &\text{ if }t<0.
  \end{eqnarray}
\end{subequations}

In summary, the stability regions of NP and SPs for different $t$ can be completely determined through (\ref{stable_NP_general}) or (\ref{stable_NP_general_t}) and (\ref{stable_SP_general}) or (\ref{stable_SP_general_t}), respectively. However, the critical boundaries of NP and SPs according to (\ref{stable_NP_general}) and (\ref{stable_SP_general}) (or (\ref{stable_NP_general_t}) and (\ref{stable_SP_general_t}), equivalently) are not exactly coincident with each other. Consequently, as shown in Fig. \ref{fig_phase} and Fig. \ref{fig_PhaseDiagram}, there arise some coexistence regions between the two, such as $p$-SP+NP and $p$-RSP+NP.

\bibliography{singleDH}

\end{document}